\documentclass[11pt,english,a4]{article}
\usepackage{lmodern}

\usepackage[T1]{fontenc}
\usepackage[latin9]{inputenc}
\usepackage[a4paper]{geometry}
\usepackage{color}
\usepackage{array}
\usepackage{multirow}
\usepackage{amsmath}
\usepackage{cancel}
\usepackage{graphicx}
\usepackage{doi}

\makeatletter

\usepackage[footnotesize]{caption}
\usepackage{booktabs}
\usepackage{graphicx} 
\usepackage[table]{xcolor} 
\usepackage{booktabs} 
\usepackage{xcolor}
\hypersetup{
    colorlinks,
    linkcolor={red!50!black},
    citecolor={blue!50!black},
    urlcolor={blue!80!black}
}
\hypersetup{colorlinks=true,linkcolor=blue}
\definecolor{grey}{rgb}{0.52, 0.52, 0.51}
\usepackage{hyphenat}
\usepackage{slashed}
\usepackage{ amssymb }
\newcommand{\PW}{\ensuremath{\text{W}\,}}
\newcommand{\pt}{\ensuremath{p_\text{T}}}
\newcommand{\MET}{\ensuremath{\slashed{E}_\text{T}}}
\newcommand{\tDM}{\ensuremath{t/\bar{t}+\text{DM}}}
\newcommand{\ttDM}{\ensuremath{t\bar{t}+\text{DM}}}

\newcommand*{\affaddr}[1]{#1} 
\newcommand*{\affmark}[1][*]{\textsuperscript{#1}}
\newcommand*{\email}[1]{\texttt{#1}}
\date{}

\@ifundefined{showcaptionsetup}{}{%
 \PassOptionsToPackage{caption=false}{subfig}}
\usepackage{subfig}
\makeatother

\usepackage{babel}
\begin{document}\sloppy


\title{Two is not always better than one:\\ Single Top Quarks and Dark Matter}

\author{%
Deborah Pinna\affmark[1], Alberto Zucchetta\affmark[1], Matthew R. Buckley\affmark[2], Florencia Canelli\affmark[1]\\
\\
\small \affaddr{\affmark[1]Department of Physics, University of Zurich,\\ \small 190 Winterthurerstrasse, 8057 Zurich, Switzerland}\\
\\
\small \affaddr{\affmark[2]Department of Physics and Astronomy, Rutgers University, \\ \small Piscataway, NJ 08854, USA}\\
\\
\footnotesize Email: \\
\footnotesize \email{deborah.pinna@cern.ch, a.zucchetta@cern.ch,} \\
\footnotesize \email{mbuckley@physics.rutgers.edu, canelli@physik.uzh.ch}\\
}

\maketitle
\begin{center}
\footnotesize (Dated: \date{\today})
\end{center}













\begin{abstract}

Dark matter interacting with the Standard Model fermions through new scalars or pseudoscalars with flavour-diagonal couplings proportional to fermion mass are well motivated theoretically, and provide a useful phenomenological model with which to interpret  experimental results. Two modes of dark matter production from these models have been considered in the existing literature: pairs of dark matter produced through top quark loops with an associated monojet in the event, and pair production of dark matter with pairs of heavy flavoured quarks (tops or bottoms). In this paper, we demonstrate that a third, previously overlooked channel yields a non-negligible contribution to LHC dark matter searches in these models. 
In spite of a generally lower production cross section at LHC when compared to the associated top-pair channel, non-flavour violating single top quark processes are kinematically favored and can significantly increase the sensitivity to these models. Including dark matter production in association with a single top quark through scalar or pseudoscalar mediators, the exclusion limit set by the LHC searches for dark matter can be improved by $30$--$90\%$, depending on the mass assumed for the mediator particle. 

\end{abstract}

\maketitle

\twocolumn

\section*{Introduction}

The nature of dark matter is one of the fundamental open questions in particle physics,
and its discovery is one of the main goals of the Large Hadron Collider (LHC) \cite{LHC_TDR}.
Gravitational evidence across a wide range of astrophysical and cosmological systems demonstrates the existence of dark matter in our Universe today \cite{Bergstrom:2012fi,Feng:2010gw}, and precision measurements from the early Universe indicates that it composes about 26\% of the Universe's energy budget \cite{Plank2015}.

Although no direct information about its properties or
non-gravitational interactions is yet available,
potential interactions between dark matter and Standard Model (SM) particles are well motivated theoretically. Assuming that dark matter was ever in thermal equilibrium in the early Universe, obtaining the observed abundance today requires dark matter to have an interaction strength with some other particle or particles at the order of the weak nuclear force or stronger \cite{Buckley:2011kk}. 
While it is possible that dark matter would be produced via other, non-thermal mechanisms, thermal production provides a tantalizing possibility for LHC discovery.
Should these interactions exist, dark matter would be produced in proton-proton
collisions at LHC and studied in detail. Collider data could access
a different range of possible interactions between dark matter and SM particles than
current direct and indirect searches \cite{lux,pamela,Bertone},
allowing an important interplay and complementarity among experiments
in the quest of discovering dark matter.

An astrophysically-viable dark matter candidate must be electrically neutral, non-interacting via the strong nuclear force, and
stable or metastable, with a decay lifetime larger than the age of the
Universe ($\sim 10^{18}$~s). 
Therefore, dark matter is invisible at the LHC, and its production at colliders can be only inferred indirectly by
a large momentum imbalance in the transverse plane
of the detector, as the dark matter recoils against visible particles. The specific nature of these visible particles is model-dependent, and given our lack of knowledge of the physics of the dark sector, it is necessary to consider all experimentally-viable possibilities.

During the LHC Run I, from 2010 to 2012, many searches for dark matter production were
performed~\cite{CMS:rwa,ATLAS:2012ky,Aad:2013oja}.
The results have been interpreted assuming contact interactions between
the dark matter and the SM sectors through effective field theories (EFTs)
\cite{Goodman:2010ku,Beltran:2010ww}. This approach is valid as long
as the energy of the interaction is such that the details of the mediator
are not resolved \cite{Busoni:2014sya,Buchmueller:2013dya}. In this
approximation, the kinematics depends only on the dark matter particle's spin and the mass
as well as the Lorentz structure of the interaction.

Considering the center-of-mass energy at LHC in Run II, it has since been realized that the assumptions underlying the EFTs would not hold at the LHC in many cases \cite{Busoni:2014sya}.
Instead, the accurate interpretation of searches for dark matter production at the LHC requires the inclusion of additional on-shell mediating particles. Integrating these mediators out (resulting in the previously-considered EFTs), often results in incorrect conclusions about the experimentally excluded regions, and may reduce the experimental sensitivity by ignoring kinematic features \cite{An:2013xka,DiFranzo:2013vra,Buchmueller:2014yoa,Papucci:2014iwa}. 
However, adding the mediator back in to the particle spectrum requires specifying the details of the interactions, which were safely ignored in the EFT approach. As an intermediate step between EFTs and a fully-specific UV theory, ``simplified models'' which have a small number of assumptions about the dark matter and contain the minimal particle content were developed to allow for accurate interpretation of LHC results~\cite{Abdallah:2014hon,Abercrombie:2015wmb}.


While the landscape of possible simplified models is large, the possibility of dark matter interactions with the SM mediated by a new scalar and/or pseudoscalar is theoretically attractive, as it can be easily accommodated in extended Higgs sectors \cite{extendedHiggs_1,extendedHiggs_2}. Given that couplings between spin-0 particles and SM fermions requires some amount of $SU(2)_L\times U(1)_Y$ breaking, it is reasonable to construct models where the scalar or pseudoscalar coupling to the SM fermions is weighted by the SM Yukawa couplings \cite{SimplifiedModel}. Assuming minimal flavour violation (MFV) \cite{Chivukula:1987py,Hall:1990ac,Buras:2000dm,D'Ambrosio:2002ex}, the discovery potential for
scalar and pseudoscalar interactions in the monojet channel (mediated by top-quark loops similar to gluon-fusion Higgs production) is significantly improved when considering
processes where the dark matter couples to massive third generation quarks
\cite{Lin:2013sca}, in particular top quarks.
This has motivated analyses searching for events in which the dark matter particles
are produced in association with a pair of top quarks
($t\bar{t}+$DM)~\cite{Aad:2014vea,CMS-PAS-EXO-16-005} or with one or two bottom quarks
($b(b)+$DM)~\cite{ATLAS-CONF-2016-086,CMS-PAS-B2G-15-007}, performed by the ATLAS and CMS collaborations with the data collected
in 2015 at $\sqrt{s}=13$ TeV. 

\begin{figure}[!htb]
\begin{centering}
\subfloat[]{\noindent \begin{centering}
\includegraphics[scale=0.2]{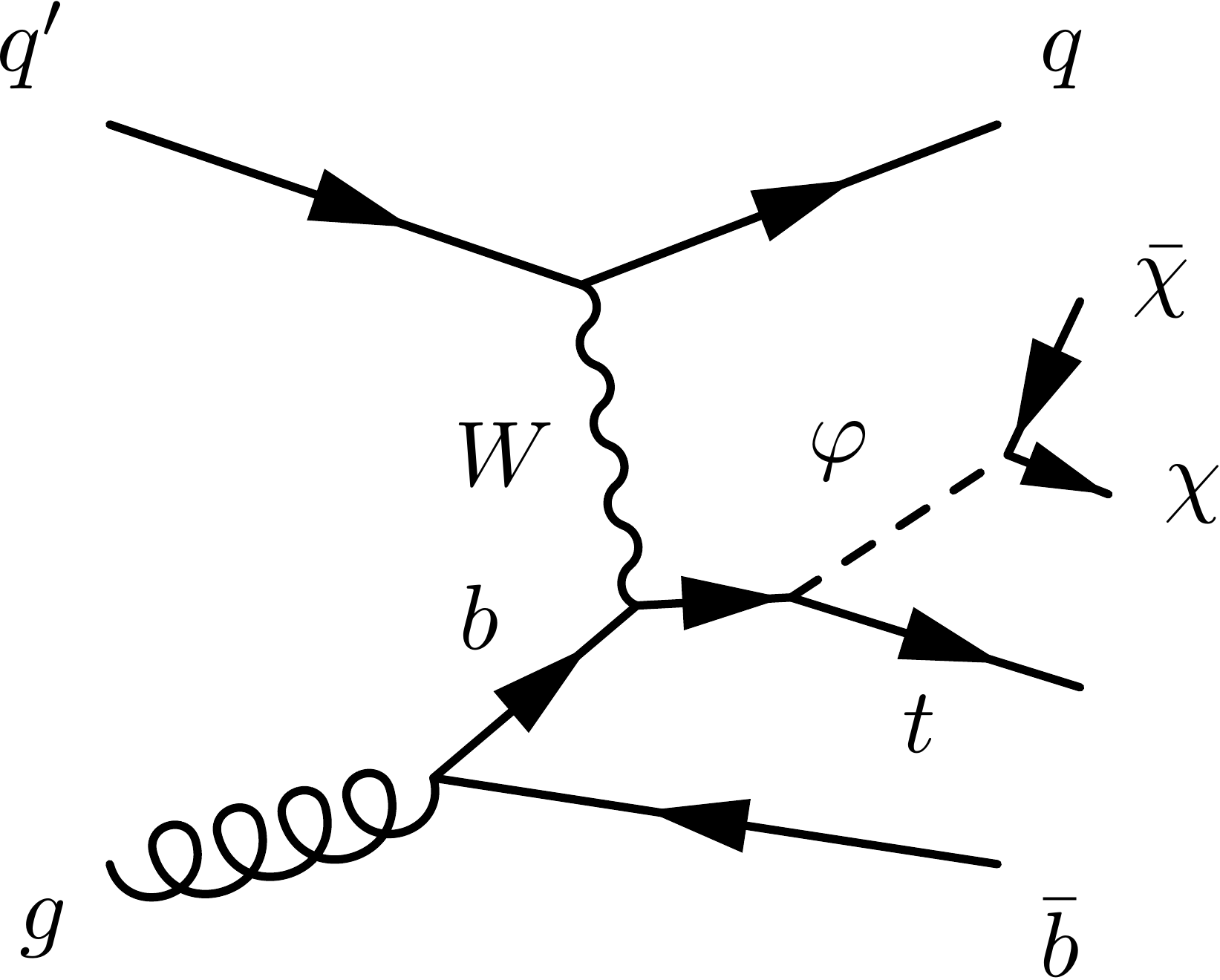}
\par\end{centering}

}\subfloat[]{\noindent \centering{}\includegraphics[scale=0.2]{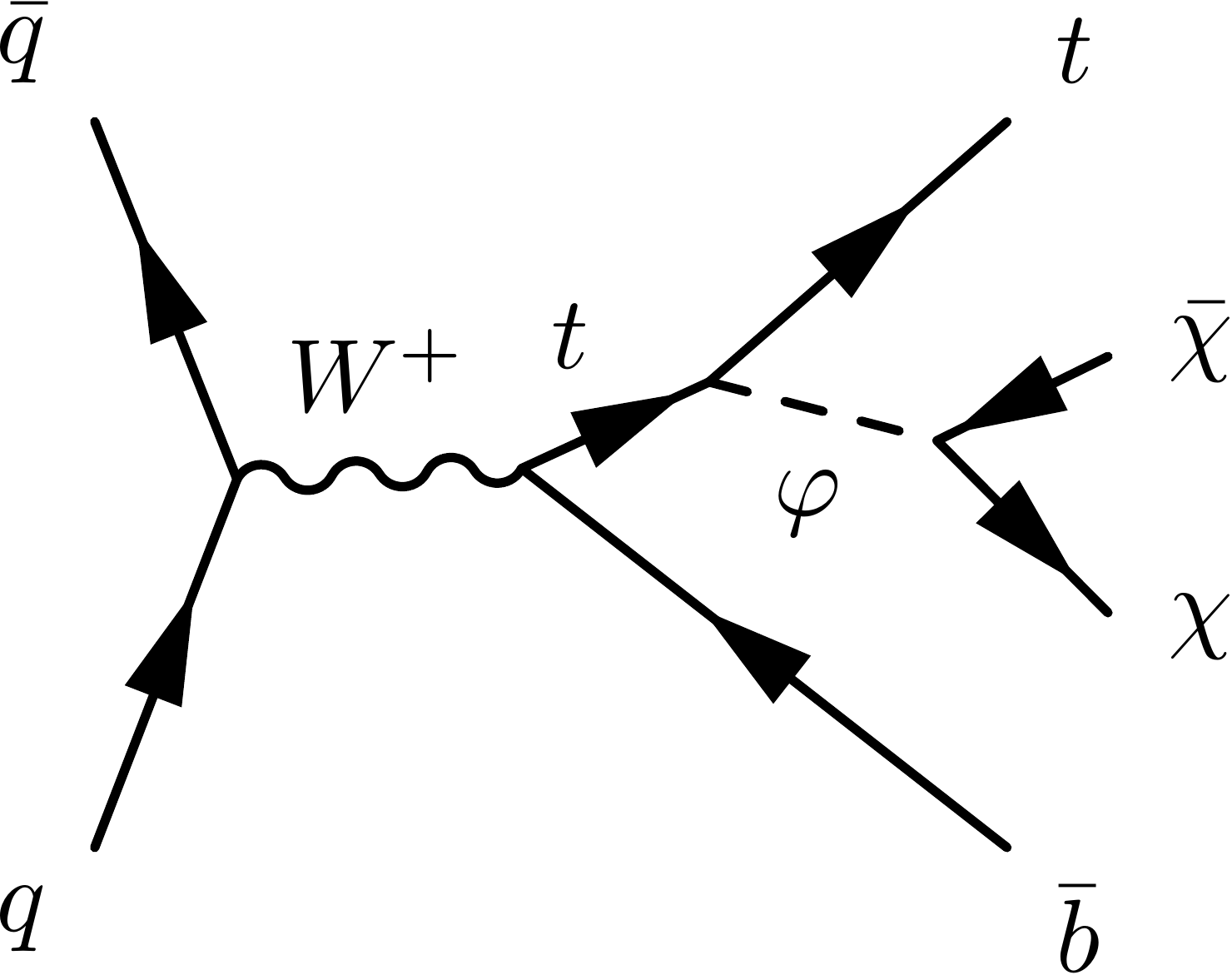}}
\par\end{centering}

\noindent \begin{centering}
\subfloat[]{\noindent \begin{centering}
\includegraphics[scale=0.2]{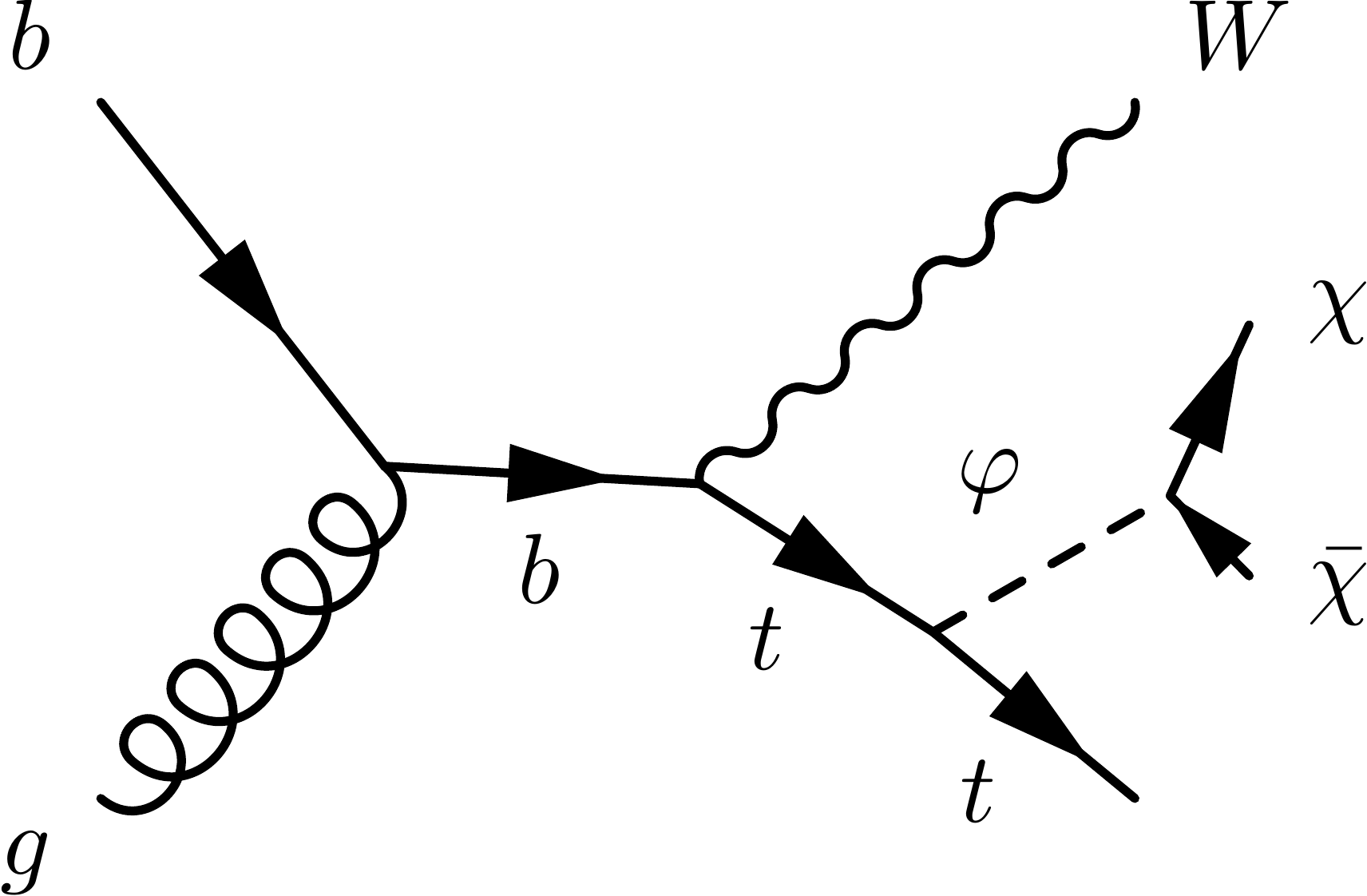}
\par\end{centering}

}\subfloat[]{\noindent \centering{}\includegraphics[scale=0.2]{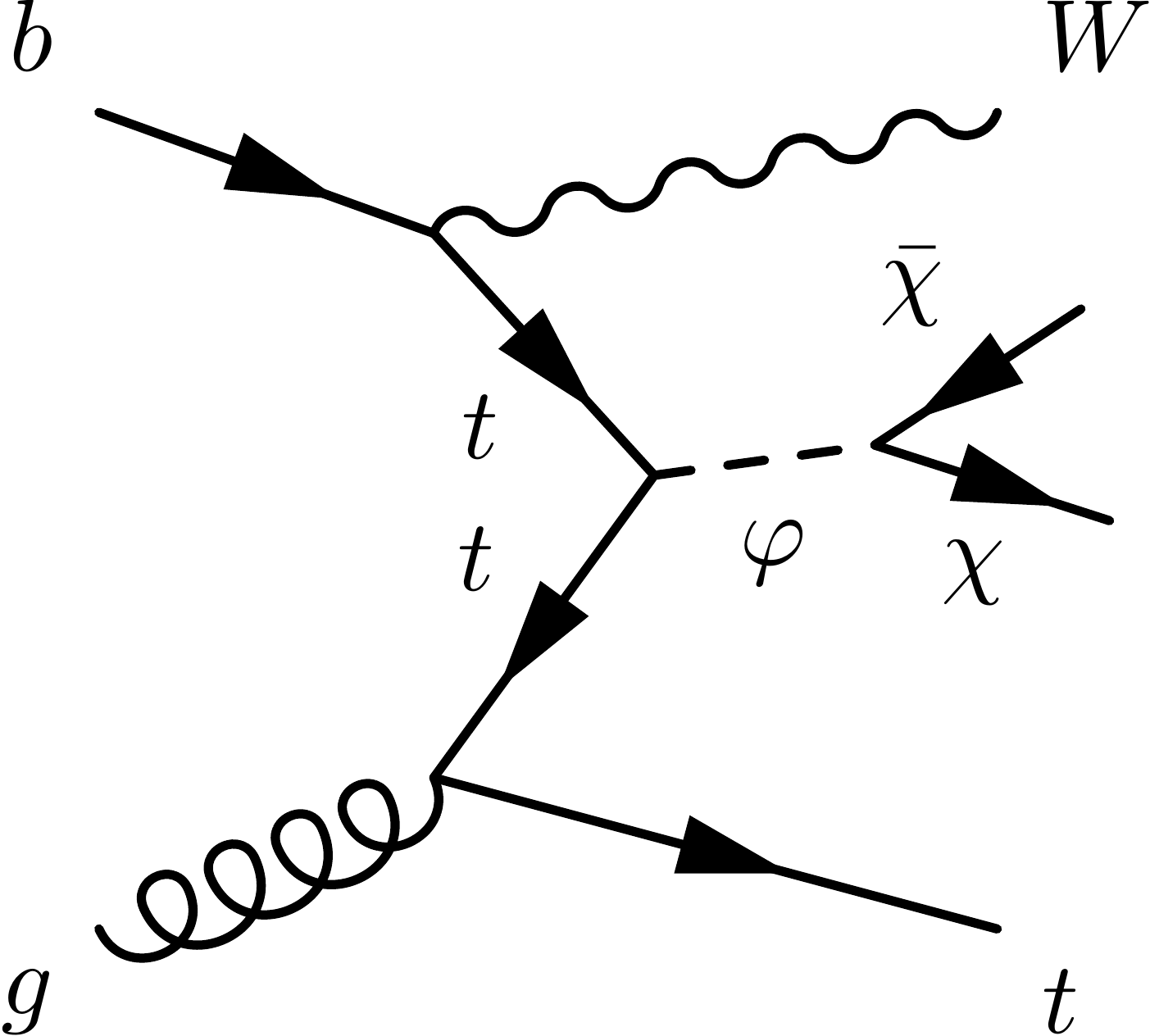}}
\par\end{centering}

\caption{\label{fig:SignalDiagram}Main production diagrams for the associated
production of dark matter with a single top at the LHC: (a) $s$--channel \PW boson production, (b) $t$--channel \PW boson production, and (c)--(d) associated $tW$ production }
\end{figure}

What has not been previously appreciated is that this same model predicts additional production mechanisms for
dark matter particles, created along with a {\it single} top quark (\tDM), rather than a pair.
The main production diagrams for this single top process are shown in Figure~\ref{fig:SignalDiagram}.
The production of the single top is obtained through processes mediated by a virtual $t$--channel or $s$--channel \PW boson (Figure~\ref{fig:SignalDiagram} (a) and (b) respectively), or through the associated production with a \PW boson (Figure~\ref{fig:SignalDiagram} (c) and (d)).
So far, final states involving a single top quark and missing energy (\MET) from dark matter particles
have been studied only considering flavour-changing neutral interactions
\cite{Khachatryan:2014uma,CMS-PAS-B2G-15-001,CMS-PAS-EXO-16-017}.

In this article we demonstrate for the first time that dark matter production in association with a single top quark, as predicted by spin-0 simplified models, yields a sizeable contribution that should be accounted for in heavy flavour searches. In spite of the generally lower cross sections, single top quark processes have a  different production mode and kinematics, resulting in overall rates comparable to top quark pair associated production, especially for a large mediator mass. As a consequence, we find that the sensitivity of the ATLAS and CMS searches can be further improved through the inclusion of this channel with respect to the \ttDM \ process alone, based on the results published by CMS in 2.2~fb$^{-1}$ of data~\cite{CMS-PAS-EXO-16-005}.

\section*{Simplified model for dark matter and single top quark production}

We assume the dark matter particles $\chi$ are Dirac fermions, with
the interaction between the SM and dark matter sectors mediated either by
a massive electrically neutral scalar $\Phi$ or a pseudo-scalar $A$ particle \cite{SimplifiedModel},
collectively referred to as $\varphi$. The Lagrangian terms of such interactions can be expressed as:
\begin{eqnarray}
\mathcal{L}_{\Phi} & \supset &g_{\chi}\Phi\bar{\chi}\chi+\frac{g_{v}\Phi}{\sqrt{2}}\sum_{f}(y_{f}\bar{f}f)\label{eq:L_scalar} \\
\mathcal{L}_{A} & \supset & ig_{\chi}A\bar{\chi}\gamma^{5}\chi+\frac{ig_{v}A}{\sqrt{2}}\sum_{f}(y_{f}\bar{f}\gamma^{5}f).\label{eq:L_pseudoscalar}
\end{eqnarray}
Here, the sum runs over the SM fermions $f$, $y_{f}=\sqrt{2}m_{f}/v$ are the Yukawa couplings with
the Higgs field vacuum expectation value $v=246$~GeV, $g_{\chi}$ is the dark matter-mediator coupling, and $g_{v}$
is the fermion-mediator coupling. For simplicity, we have assume a universal value of $g_v$ for all
fermion flavours.
Under the MFV assumption, this simplified model has a minimal set of four free parameters:
($m_{\chi}, \ m_{\varphi}, \ g_{\chi},\ g_{v}$).

It is most convenient to present experimental results in terms of the product of the visible and invisible couplings $g_\chi$, $g_v$ as a function of the two masses $(m_\varphi,m_\chi)$. However, while the overall production rate times decay is set by this product squared, it also depends on the width $\Gamma_\varphi$, set by $g_\chi^2$ and $g_v^2$ separately:
\begin{equation}
\begin{aligned}\Gamma_{\varphi}= & \frac{g_\chi^2m_{\varphi}}{8\pi}\left(1-\frac{4m_{\chi}^{2}}{m_{\varphi}^{2}}\right)^{n/2}\\
 & +\sum_{f}\frac{g_v^2y_{f}^{2}m_{\varphi}}{16\pi}\left(1-\frac{4m_{f}^{2}}{m_{\varphi}^{2}}\right)^{n/2}
\end{aligned}
\label{eq:MinWidth}
\end{equation}
where $n=3$ for scalars and $n=1$ for pseudo-scalars interactions
respectively. One can make a number of simplifying assumptions to specify the mediator width in terms of $g_\chi$, $g_v$, $m_\varphi$, and $m_\chi$. In this paper, we assume that $g_\chi = g_v = 1$. Note that this makes a model-specific assumption for the cross sections and branching ratios of the mediator to dark matter and SM fermions. This should be kept in mind when comparing the experimental sensitivity of various channels; though it does not affect the particular comparisons we make in this paper. More details about this simplified model are described in Ref.~\cite{Abercrombie:2015wmb,SimplifiedModel}. 

\begin{figure*}
\noindent \begin{centering}
\subfloat[]{\noindent \begin{centering}
\includegraphics[scale=0.38]{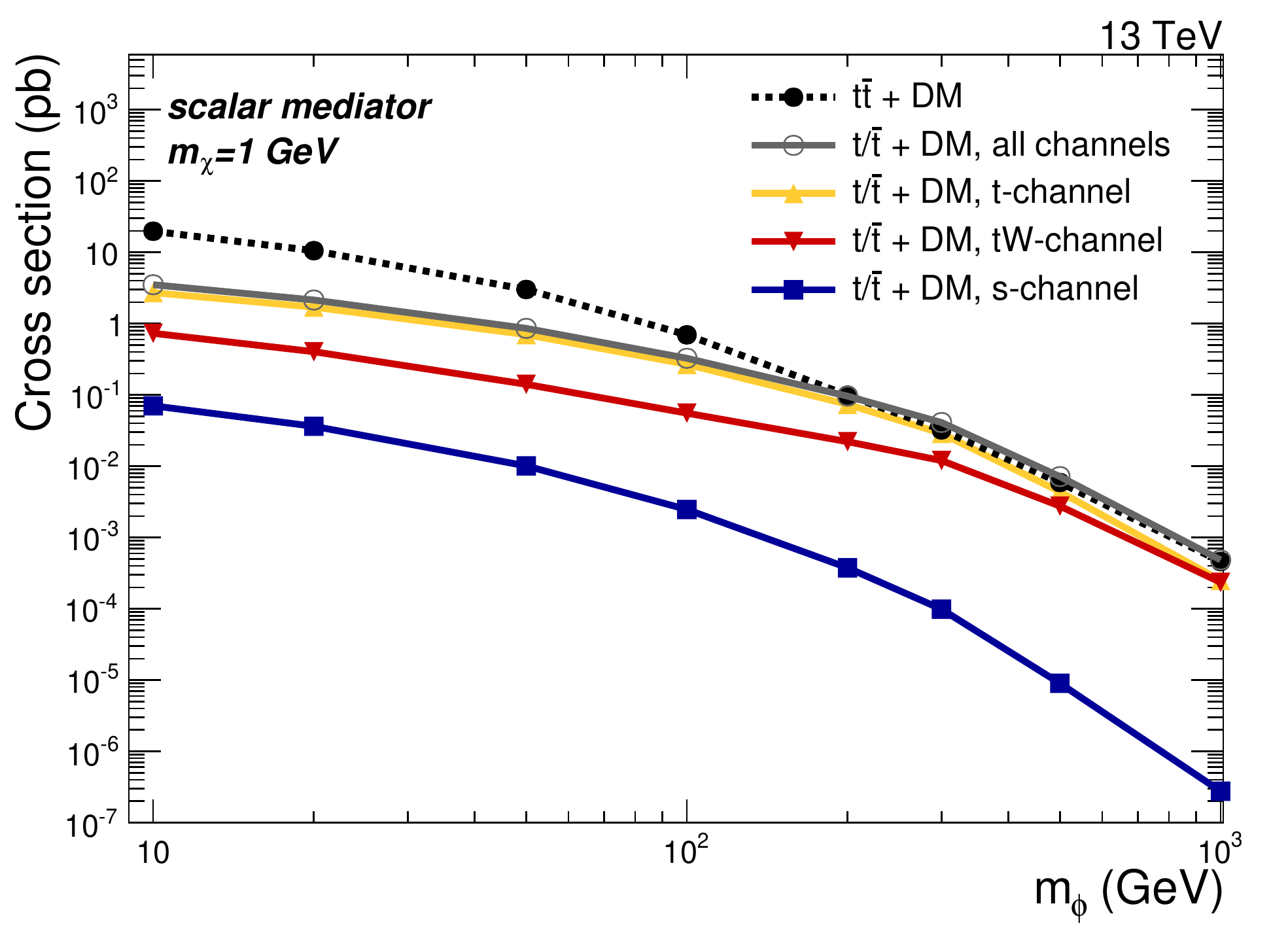}
\par\end{centering}
}
\subfloat[]{\noindent \centering{}\includegraphics[scale=0.38]{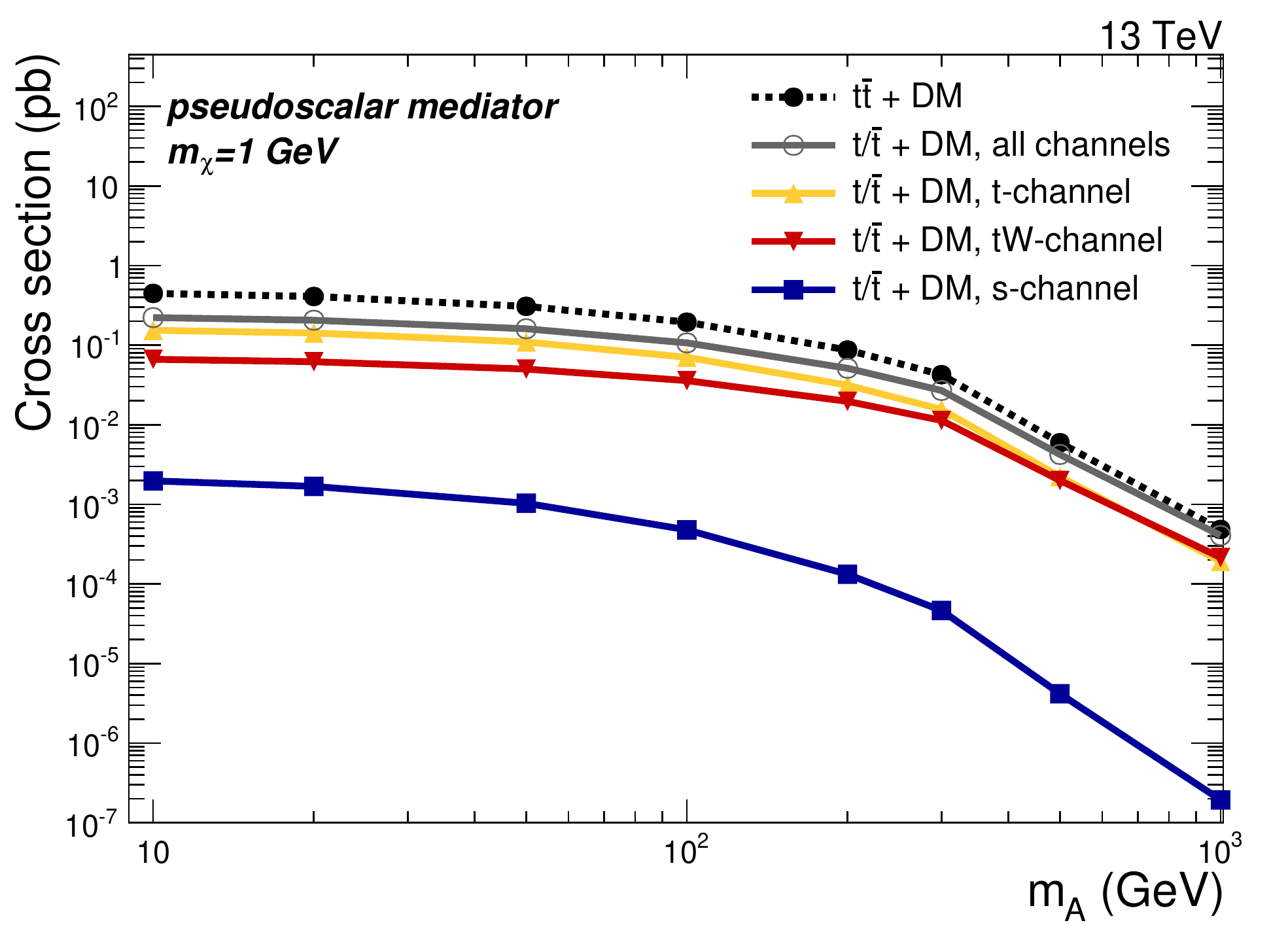}}
\caption{\label{fig:xsecPlot} Cross sections of the \tDM\, and \ttDM\, processes for the scalar (a) or pseudoscalar (b) hypothesis assuming different mediator masses $m_\varphi$ and $m_\chi=1$ GeV. The \tDM\, processes are split by production mode ($t$--, $s$--, and $tW$ channels). The sum of the three channels is also shown. \label{fig:xs}}
\par\end{centering}
\end{figure*}

The signal processes are simulated at tree level with the {\sc MG5\_aMC@NLO} v2.4.3 generator~\cite{MG_amNLO}, without additional quarks or gluons in the parton-level interaction. The {\sc Pythia} v6.428 software~\cite{Pythia6} is used for showering and hadronization.
The \ttDM\, and \tDM\, $tW$-channel
processes are produced in the massless $b$-quark hypothesis
(5-flavour scheme), while the \tDM\, $t$-- and $s$--channel processes
are generated considering the $b$ quark as a massive particle (4-flavour scheme).
Cross sections at generator level are calculated in 5-flavour scheme
for all processes 
 and are reported for different
mediator and dark matter mass hypotheses in Table~\ref{tab:xsec_sc}. The same values are presented graphically in Figure~\ref{fig:xsecPlot} for the hypothesis $m_\chi=1$ GeV.
For each mass point and process, 200k events have been generated, except for \tDM\, $t$--channel for which this number is increased to 500k.

\begin{table*}[!htb]
\resizebox{\textwidth}{!}{
\begin{tabular}{cc||c|cccc}
\toprule 
&\multirow{2}{*}{$m_\chi,\ m_\varphi$ (GeV)} & \multirow{2}{*}{\ttDM ($\mathrm{pb}$)} & \multicolumn{4}{c}{\tDM ($\mathrm{pb}$)}\\
& &  & $t$--channel & $tW$--channel & $s$--channel & total \tDM\\
\midrule
\multirow{10}{*}{\rotatebox[origin=c]{90}{scalar}} & $m_\chi=1,\ m_\Phi=10$ 	& $19.76\pm 0.01$  				&$(27.18\pm 0.05) \cdot 10^{-1}$		&$(73.25\pm 0.06)\cdot 10^{-2}$		&$(7.03\pm 0.01) \cdot 10^{-2}$		&$(35.20\pm 0.05) \cdot 10^{-1}$\\
 & $m_\chi=1,\ m_\Phi=20$ 	& $(10.55\pm 0.01)$  			&$(17.03\pm 0.03) \cdot 10^{-1}$		&$(40.44\pm 0.03)\cdot 10^{-2}$		&$(36.29\pm 0.06) \cdot 10^{-3}$		&$(21.43\pm 0.03) \cdot 10^{-1}$\\
 & $m_\chi=1,\ m_\Phi=50$ 	& $(30.06\pm 0.02) \cdot 10^{-1}$  	& $(7.00\pm 0.01) \cdot 10^{-1}$		&$(14.09   \pm 0.01)\cdot 10^{-2}$		&$(10.10\pm  0.02) \cdot 10^{-3}$		&$(8.51\pm 0.01) \cdot 10^{-1}$\\
 & $m_\chi=1,\ m_\Phi=100$ 	& $(69.60\pm 0.04) \cdot 10^{-2}$  	&$(26.83\pm 0.04)	 \cdot 10^{-2}$		&$(55.49 \pm 0.04) \cdot 10^{-3}$		&$(24.74\pm   0.03) \cdot 10^{-4}$		&$(32.62\pm 0.04) \cdot 10^{-2}$\\
 & $m_\chi=1,\ m_\Phi=200$ 	& $(99.16\pm 0.07)  \cdot 10^{-3}$  	&$(7.37\pm    0.01) \cdot 10^{-2}$		&$(22.15 \pm 0.02) \cdot 10^{-3}$		&$(37.6\pm     0.05) \cdot 10^{-5}$		&$(9.62\pm 0.01) \cdot 10^{-2}$\\
 & $m_\chi=1,\ m_\Phi=300$ 	& $(32.21\pm 0.02)  \cdot 10^{-3}$  	&$(28.88\pm  0.05) \cdot 10^{-3}$		&$(12.04 \pm 0.01) \cdot 10^{-3}$		&$(9.87\pm   0.02) \cdot 10^{-5}$		&$(41.02\pm 0.005) \cdot 10^{-3}$\\
 & $m_\chi=1,\ m_\Phi=500$ 	& $(59.00\pm 0.06) \cdot 10^{-4}$ 	&$(43.85\pm 0.08) \cdot 10^{-4}$		&$(27.61  \pm 0.02) \cdot 10^{-4}$	&$(9.01\pm 0.01) \cdot 10^{-6}$		&$(71.55\pm  0.08) \cdot 10^{-4}$\\
 & $m_\chi=1,\ m_\Phi=1000$ 	& $(46.03\pm 0.05) \cdot 10^{-5}$  	&$(24.99\pm 0.03) \cdot 10^{-5}$		&$(23.46\pm  0.02) \cdot 10^{-5}$		&$(27.64\pm 0.04) \cdot 10^{-8}$		&$(48.48\pm0.04) \cdot 10^{-5}$\\
 & $m_\chi=10,\ m_\Phi=10$ 	& $(96.42\pm 0.07) \cdot 10^{-3}$  	&$(23.13\pm 0.04) \cdot 10^{-3}$		&$(50.44  \pm 0.04) \cdot 10^{-4}$	&$(32.83\pm 0.07) \cdot 10^{-5}$		&$(28.51\pm 0.04) \cdot 10^{-3}$\\
 & $m_\chi=50,\ m_\Phi=300$ 	& $(31.86\pm 0.03) \cdot 10^{-3}$ 	&$(28.73\pm 0.04) \cdot 10^{-3}$		&$(120.34   \pm 0.09) \cdot 10^{-}4$	&$(9.74\pm 0.02) \cdot 10^{-5}$		&$(40.86\pm 0.04) \cdot 10^{-3}$\\
\midrule 
\multirow{10}{*}{\rotatebox[origin=c]{90}{pseudoscalar}} & $m_\chi=1,\ m_A=10$ 		& $(44.63\pm  0.03) \cdot 10^{-2}$ 	&$(15.34\pm  0.02) \cdot 10^{-2}$ 		& $(66.36\pm 0.04) \cdot 10^{-3}$ &  	    $(19.69\pm 0.04) \cdot 10^{-4}$ 		&$(22.19\pm  0.02) \cdot 10^{-2}$ \\
& $m_\chi=1,\ m_A=20$ 		& $(40.80\pm 0.03) \cdot 10^{-2}$	& $(14.19\pm 0.02) \cdot 10^{-2}$ 		& $(61.82\pm  0.04) \cdot 10^{-3}$ 		&  $(16.78\pm 0.03) \cdot 10^{-4}$ 		&$(20.53\pm 0.02) \cdot 10^{-2}$ \\
& $m_\chi=1,\ m_A=50$ 		& $(30.72\pm  0.03) \cdot 10^{-2}$ 	& $(10.94\pm 0.02) \cdot 10^{-2}$ 		& $(50.00\pm 0.04) \cdot 10^{-3}$ 		&  $(10.30\pm 0.02) \cdot 10^{-4}$ 		&$(16.04\pm 0.02) \cdot 10^{-2}$ \\
& $m_\chi=1,\ m_A=100$ 		& $(19.41\pm 0.02) \cdot 10^{-2}$  	& $(7.04\pm 0.01) \cdot 10^{-2}$ 		& $(35.79\pm 0.03) \cdot 10^{-3}$		&  $(47.79\pm 0.09) \cdot 10^{-5}$	 	&$(10.66\pm  0.01) \cdot 10^{-2}$ \\
& $m_\chi=1,\ m_A=200$ 		& $(86.78\pm 0.08) \cdot 10^{-3}$ 	& $(31.39\pm 0.05) \cdot 10^{-3}$ 		& $(19.65\pm 0.02) \cdot 10^{-3}$ 		&  $(13.20\pm 0.02) \cdot 10^{-5}$ 		&$(51.17\pm  0.05) \cdot 10^{-3}$ \\
& $m_\chi=1,\ m_A=300$		& $(42.50\pm 0.04) \cdot 10^{-3}$ 	& $(15.55\pm 0.02) \cdot 10^{-3}$ 		& $(11.33\pm  0.01) \cdot 10^{-3}$ 		&  $(46.28\pm 0.08) \cdot 10^{-6}$ 		&$(26.92\pm 0.03) \cdot 10^{-3}$ \\
& $m_\chi=1,\ m_A=500$ 		& $(59.43\pm 0.06) \cdot 10^{-4}$	& $(22.27\pm 0.04) \cdot 10^{-4}$ 		& $(19.96\pm 0.02) \cdot 10^{-4}$ 		&  $(41.72\pm 0.07) \cdot 10^{-7}$ 		&$(42.27\pm  0.05) \cdot 10^{-4}$ \\ 
& $m_\chi=1,\ m_A=1000$		& $(48.33\pm 0.05) \cdot 10^{-5}$	& $(19.09\pm 0.03) \cdot 10^{-5}$ 		& $(21.44\pm 0.02) \cdot 10^{-5}$ 		&  $(19.31\pm 0.03) \cdot 10^{-8}$ 		&$(40.56\pm 0.03) \cdot 10^{-5}$ \\
& $m_\chi=10,\ m_A=10$ 		& $(15.28\pm 0.02) \cdot 10^{-3}$ 	& $(5.45\pm   0.01) \cdot 10^{-3}$  		& $(26.74\pm 0.02) \cdot 10^{-4}$ 		&  $(47.17\pm 0.09) \cdot 10^{-6}$ 		&$(8.17\pm  0.01) \cdot 10^{-3}$ \\
& $m_\chi=50,\ m_A=300$ 	& $(42.43\pm 0.04) \cdot 10^{-3}$ 	& $(15.54\pm 0.03) \cdot 10^{-3}$		& $(11.34\pm  0.01) \cdot 10^{-3}$ 		&  $(46.19\pm 0.08) \cdot 10^{-6}$ 		&$(26.93\pm  0.03) \cdot 10^{-3}$ \\
\bottomrule 
\end{tabular}}\caption{\label{tab:xsec_sc} Cross sections of the \tDM\, and \ttDM\, processes, for different mediator and dark matter masses. Both the scalar and pseudoscalar hypotheses are considered. The \tDM\, processes are split by production mode ($t$--, $s$--, and $tW$ channels). The sum of the three is also provided. These cross sections are also shown in Figure~\ref{fig:xs} for $m_\chi = 1$~GeV.}
\end{table*}

The response of the CMS detector is simulated with {\sc Delphes} v3.3.3~\cite{Delphes}.
Reconstruction and identification efficiencies for electrons
and muons are taken from Ref.~\cite{electron} and Ref.~\cite{muon} respectively.
Charged and neutral particles are clustered into jets using the anti-$k_\text{t}$ algorithm~\cite{anti-kt} with radius parameter $R=0.4$, implemented in the {\sc FastJet} package~\cite{Fastjet}. Jets originating from bottom-flavoured hadrons are identified with an algorithm $60\%$ efficient for $b$-quarks, and with a mis-identification rate of $8\%$
for $c$ quarks and $0.1\%$ for light quarks~\cite{btag}.
In order to reproduce the conditions during the 2015 data-taking,
an average of 11 additional parton interactions are simulated for every bunch crossing.

\section*{Selection criteria and kinematics}

In the \ttDM\, CMS analysis~\cite{CMS-PAS-EXO-16-005}, 
reconstructed events are split in two exclusive categories, depending on the number of leptons in the final state. As the search assumes the pair production of top quarks in association with dark matter, the zero-lepton final state corresponds to the hadronic decay of both tops, while the one-lepton state comes from semi-leptonic decay. The selections applied are different in the two cases, due to the different background composition and trigger requirements.

In the hadronic channel, events with isolated leptons and transverse momentum $\pt > 10$ GeV are rejected. A minimum of four hadronic jets with $\pt>30$ GeV are required, and at least two of these must satisfy the $b$-tagging selection. In order to reject multijet events, the minimum azimuthal separation ($\Delta \phi$) between the missing energy and the six highest-$\pt$ jets in the event must be larger than $1.0$. The signal-enriched region is characterized by a large amount of missing energy ($\MET > 200$ GeV) reconstructed in the detector.

The semileptonic channel requires exactly one isolated lepton ($\ell$, electron or muon) with $\pt > 30$ GeV and pseudorapidity $|\eta| < 2.1$, and no other isolated leptons with $\pt > 10$ GeV. Due to the leptonic trigger, the \MET\, threshold is lowered to 160 GeV. The minimum jet multiplicity is also reduced to three or more jets, with 
at least of one $b$-tagged jet. 
Selections on the \PW boson transverse mass, ($m^W_T = \sqrt{2 \MET E_{\text{T}}^\ell \cdot [1-\cos \Delta \phi {(\ell,\MET)} ]}$),
and the $m_{T2}^{W}$ variable~\cite{mt2w} are used to discriminate against SM \PW boson production and dileptonic $t\bar{t}$ backgrounds, respectively.

\begin{figure*}
\noindent \begin{centering}
\subfloat[]{\noindent \begin{centering}
\includegraphics[scale=0.38]{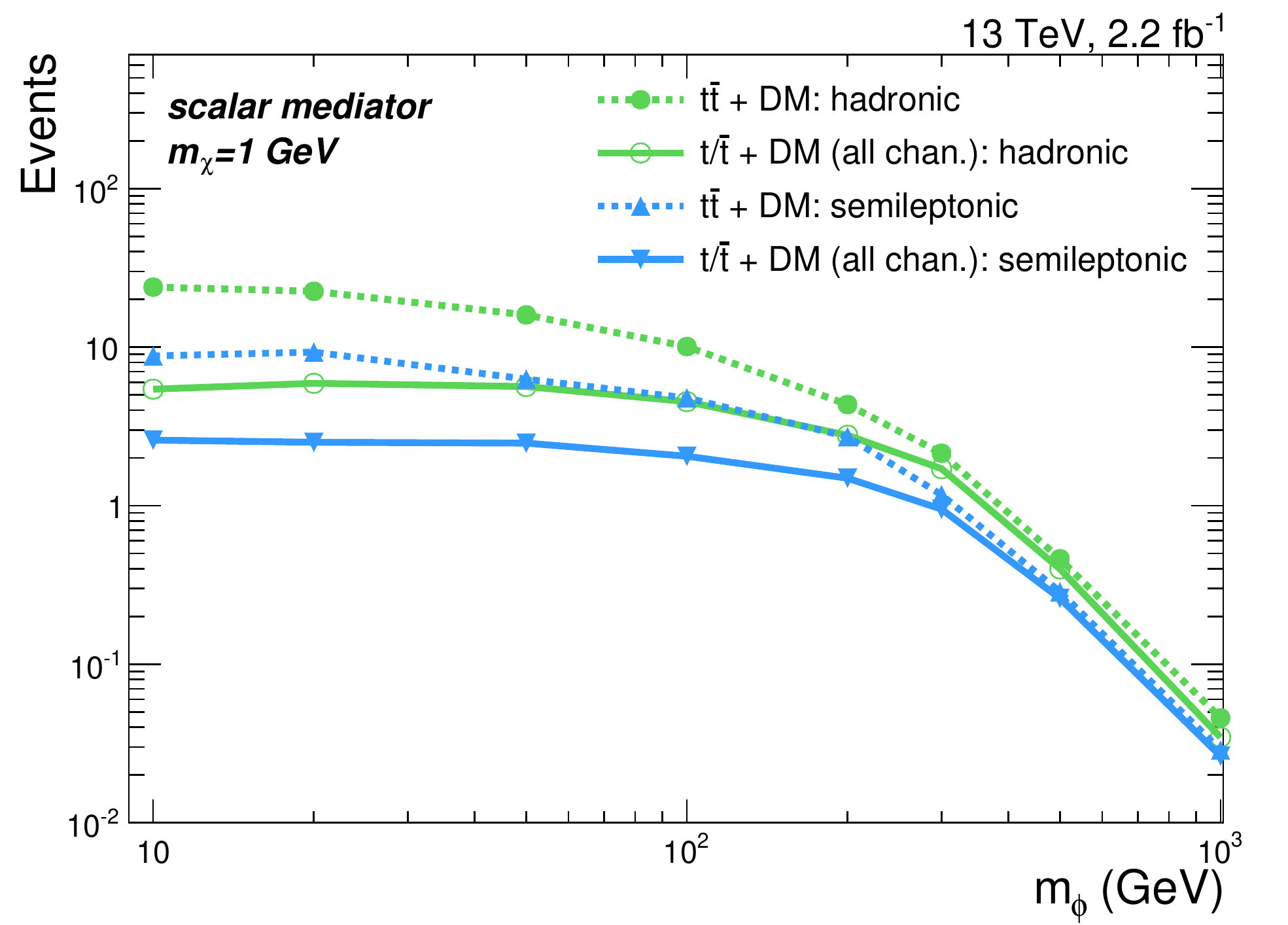}
\par\end{centering}
}
\subfloat[]{\noindent \centering{}\includegraphics[scale=0.38]{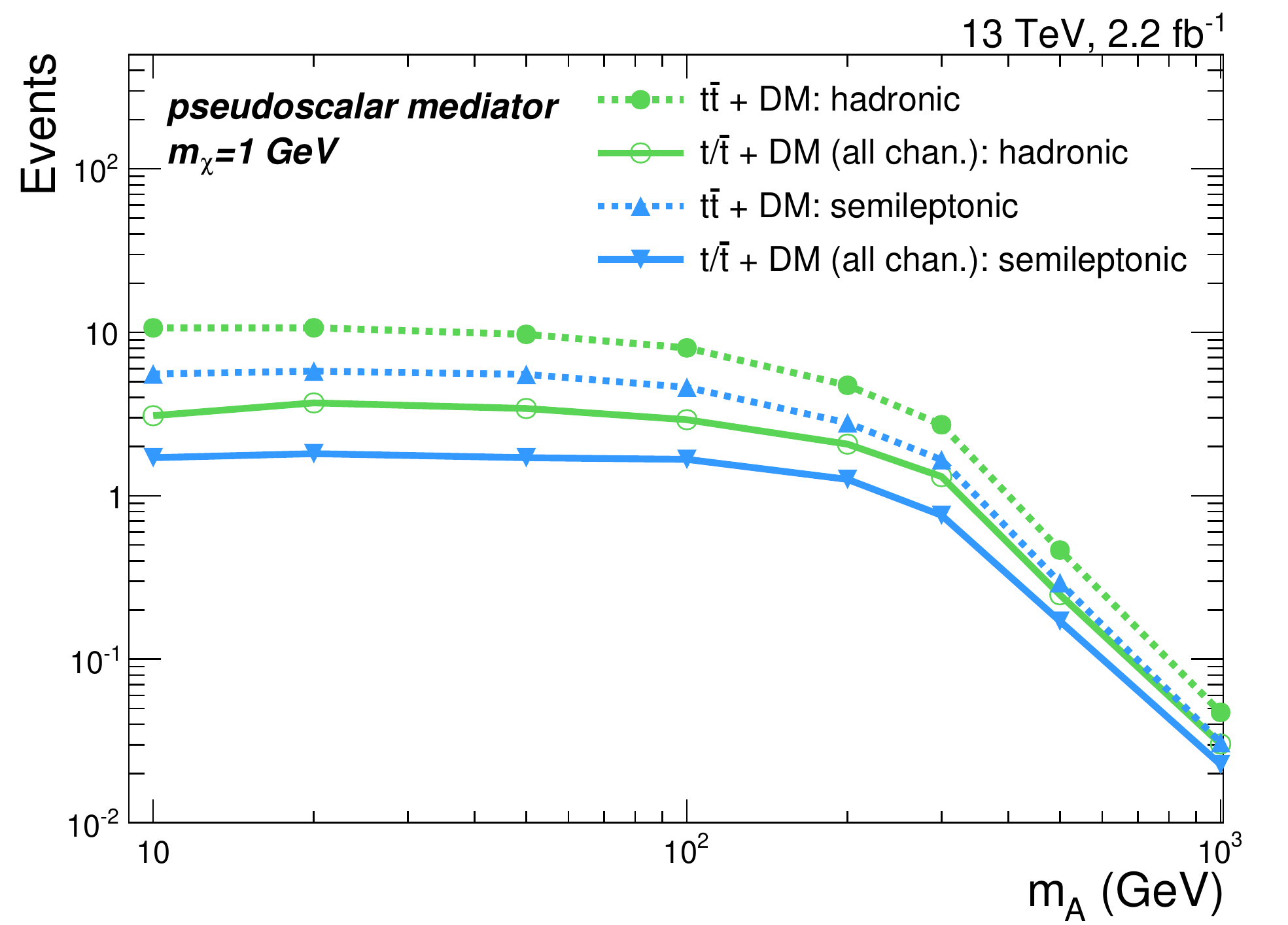}}
\caption{\label{fig:yieldPlot} Expected signal events for scalar (a) and pseudoscalar (b) mediators for various $m_\varphi$ mass choices and $m_\chi=1$ GeV. The numerical values are referred to an integrated luminosity of $2.2~\mbox{fb}^{-1}$, and are separated by process (\ttDM\, or the sum of all \tDM\, production channels) and final state (hadronic or semileptonic).}
\par\end{centering}
\end{figure*}

The list of selection criteria relative to the two categories
are presented in Table \ref{tab:Selection}, and the number of signal events corresponding to the 2015 dataset are reported in Table~\ref{tab:yields_SR}.
The signal yields are also presented graphically in Figure~\ref{fig:yieldPlot} for different mediator mass hypotheses and $m_\chi=1$ GeV.

\begin{table}
\resizebox{\columnwidth}{!}{
\begin{tabular}{c c c}
\toprule 
\multirow{2}{*}{selection} & \multicolumn{2}{c}{channel}\\
 & hadronic & semileptonic\\
\midrule
\MET & $>200$ GeV & $>160$ GeV\\
number of jets & $\geq 4$ & $\geq 3$\\
number of b-jets & $\geq 2$ & $\geq 1$\\
number leptons & veto & $1$\\
$p_{T}^\ell$ & $>10$ GeV & $>30$ GeV\\
\multirow{2}{*}{$|\eta^\ell|$} & $<2.5$ (electrons) & $<2.1$ (electrons)\\
 & $<2.4$ (muons) & $<2.1$ (muons)\\
$\Delta\phi(j_{i},\MET)$ & $>1;\ i=1,..,6$ & $>1.2;\ i=1,2$\\
$m_{T}^{W}$ & $-$ & $>160$ GeV\\
$m_{T2}^{W}$ & $-$ & $>200$ GeV\\
\bottomrule 
\end{tabular}}
\caption{\label{tab:Selection} Event selections for the hadronic and semileptonic channels, as applied in the CMS analysis~\cite{CMS-PAS-EXO-16-005}.}
\end{table}

\begin{table*}[!htb]
\resizebox{\textwidth}{!}{
\begin{tabular}{cc cc cc}
\toprule 
&\multirow{2}{*}{$m_\chi,\ m_{\varphi}$ (GeV)} & \multicolumn{2}{c}{\ttDM } & \multicolumn{2}{c}{\tDM }\\
&  &  hadronic & semileptonic & hadronic & semileptonic \\
\midrule
\multirow{10}{*}{\rotatebox[origin=c]{90}{scalar}} 
& $m_\chi=1,\ m_\Phi=10$ &       	$23.9\pm 2.3$  &				$8.8 \pm 1.8$&							$5.43\pm 0.45$&					$2.59 \pm 0.33$\\
 & $m_\chi=1,\ m_\Phi=20$ & 	$22.5\pm 1.6$  &				$9.3 \pm 1.4$&			 				$5.91\pm 0.48$&					$2.51 \pm 0.33$\\	
 & $m_\chi=1,\ m_\Phi=50$ & 	$15.96\pm  0.90$  & 				$6.28 \pm 0.73$&							$5.64 \pm 0.23$&					$2.48 \pm 0.17$\\
 & $m_\chi=1,\ m_\Phi=100$ & 	$10.08\pm 0.42$  &				$4.78 \pm 0.38$&							$4.53\pm 0.17$&					$2.05 \pm 0.12$\\
 & $m_\chi=1,\ m_\Phi=200$ & 	$4.34\pm 0.07)$  & 				$2.72 \pm 0.07$&							$2.78\pm 0.07$&					$1.49\pm 0.05$\\
 & $m_\chi=1,\ m_\Phi=300$ & 	$2.14\pm 0.03)$  &				$1.17 \pm 0.03$ &							$1.71\pm 0.03$&					$0.95 \pm 0.02$\\
 & $m_\chi=1,\ m_\Phi=500$ & 	$(4.62\pm 0.05) \cdot 10^{-1}$ &	$(2.86 \pm 0.06) \cdot 10^{-1}$&				$(3.99\pm 0.05) \cdot 10^{-1}$&		$(2.60 \pm 0.05) \cdot 10^{-1}$\\
 & $m_\chi=1,\ m_\Phi=1000$ & 	$(4.58\pm 0.05) \cdot 10^{-2}$&	$(2.87 \pm 0.05) \cdot 10^{-2}$&				$(3.46\pm 0.04) \cdot 10^{-2}$&		$(2.63\pm 0.04) \cdot 10^{-2}$\\
 & $m_\chi=10,\ m_\Phi=10$ & 	$0.61\pm 0.03 $  & 				$0.28 \pm 0.02$&							$(2.56\pm 0.09) \cdot 10^{-1}$&		$(2.84 \pm 0.23) \cdot 10^{-1}$\\
 & $m_\chi=50,\ m_\Phi=300$ & 	$2.21\pm 0.03 $ &				$1.16 \pm 0.03$&							$1.74\pm 0.07$&					$0.87 \pm 0.04$\\
\midrule
\multirow{10}{*}{\rotatebox[origin=c]{90}{pseudoscalar}} 
& $m_\chi=1,\ m_A=10$ & 		$10.65\pm 0.23$ &				$5.56 \pm 0.22$ & 							$3.09\pm 0.05$ &  					$1.71 \pm 0.05$ \\
& $m_\chi=1,\ m_A=20$ &  		$10.66\pm 0.22$ & 				$5.79 \pm 0.21$ & 							$3.70\pm 0.11$ &  					$1.81 \pm 0.08$ \\
& $m_\chi=1,\ m_A=50$ & 		$9.72\pm 0.18$ & 				$5.53 \pm 0.18$ & 							$3.42\pm 0.08$ & 					$1.71 \pm 0.06$ \\
& $m_\chi=1,\ m_A=100$ & 		$8.03\pm 0.13$  & 				$4.61 \pm 0.13$ &  							$2.92\pm 0.06$&  					$1.67 \pm 0.05$ \\
& $m_\chi=1,\ m_A=200$ & 		$4.74\pm 0.07$ & 				$2.79 \pm 0.07$ & 							$2.07\pm 0.04$ &  					$1.26 \pm 0.03$\\
& $m_\chi=1,\ m_A=300$ & 		$2.72\pm 0.04$ & 				$1.66 \pm 0.04$ & 							$1.31\pm 0.02$ & 					$0.76 \pm 0.01$ \\
& $m_\chi=1,\ m_A=500$ & 		$(4.65\pm 0.06) \cdot 10^{-1}$ & 	$(2.94 \pm 0.06) \cdot 10^{-1}$ &				$(2.48\pm 0.03) \cdot 10^{-1}$ &  $(1.71 \pm 0.03) \cdot 10^{-1}$\\ 
& $m_\chi=1,\ m_A=1000$ & 	$(4.73\pm 0.05) \cdot 10^{-2}$ & 	$(3.09 \pm 0.05) \cdot 10^{-2}$ &				$(3.03\pm 0.03) \cdot 10^{-2}$ & 	$(2.27 \pm 0.03) \cdot 10^{-2}$ \\
& $m_\chi=10,\ m_A=10$ & 		$0.54\pm 0.01$ & 				$0.31 \pm 0.01$  &	 						$(2.05\pm 0.06) \cdot 10^{-1}$ & 		$(1.11 \pm 0.05) \cdot 10^{-1}$\\
& $m_\chi=50,\ m_A=300$ & 	$2.78\pm 0.04$ &  				$1.68 \pm 0.04$ & 							$1.25\pm 0.04$ &  					$0.79 \pm 0.03$\\
\midrule
\multicolumn{2}{c}{Background} & $324 \pm 22$ & $47.1 \pm 3.5$ & $324 \pm 22$ & $47.1 \pm 3.5$ \\
\bottomrule 
\end{tabular}}\caption{\label{tab:yields_SR} Signal and background events for scalar and pseudoscalar mediators for various $m_\chi,\ m_\varphi$ mass choices. The numerical values are referred to an integrated luminosity of $2.2~\mbox{fb}^{-1}$, and are separated by process (\ttDM\, or the sum of all \tDM\, production channels) and final state (hadronic or semileptonic). The background events are taken from the CMS analysis~\cite{CMS-PAS-EXO-16-005}.}
\end{table*}

The detector parametrization is validated by comparing the \ttDM\, signal yields estimated with {\sc Delphes} with the values from the
CMS search~\cite{CMS-PAS-EXO-16-005}.
The \ttDM\, samples used for the validation are produced assuming
a dark matter mass of 1 GeV and a mediator mass of 100 GeV, for both the scalar
and pseudoscalar hypothesis. An additional sample with 1 GeV dark matter particle
and a mediator mass of 10 GeV is considered for the scalar interaction.
The yields are reported in Table \ref{tab:Yields}, and are found to be compatible with the experimental result within the statistical uncertainty.


\begin{table}
\noindent \centering{}\resizebox{\columnwidth}{!}{
\begin{tabular}{c cc cc}
\toprule 
\multirow{2}{*}{$m_\chi,\ m_\varphi$ (GeV)} & \multicolumn{2}{c}{hadronic} & \multicolumn{2}{c}{semileptonic}\\
 & CMS & {\sc Delphes} & CMS & {\sc Delphes}\\
\midrule 
$m_\chi=1,\ m_{\phi}=10$ & $20\pm 12$ & $23.9\pm 2.3$ & $9.1 \pm 4.3$ & $8.8 \pm 1.8$\\
$m_\chi=1,\ m_{\phi}=100$ & $10.0\pm 3.0$ & $10.1\pm 0.42$ & $4.64\pm 0.56$ & $4.78 \pm 0.38$ \\
$m_\chi=1,\ m_{A}=100$ & $8.5\pm 1.4$ & $8.03\pm 0.13$ & $4.36\pm 0.29$ & $4.61 \pm 0.13$\\
\bottomrule 
\end{tabular}}
\caption{\label{tab:Yields} Comparison of the number of \ttDM\, events between the CMS analysis~\cite{CMS-PAS-EXO-16-005} and the {\sc Delphes} parametrization. The uncertainty reported is statistical only.}
\end{table}

\begin{figure*}
\noindent \begin{centering}
\subfloat[]{\noindent \begin{centering}
\includegraphics[scale=0.26]{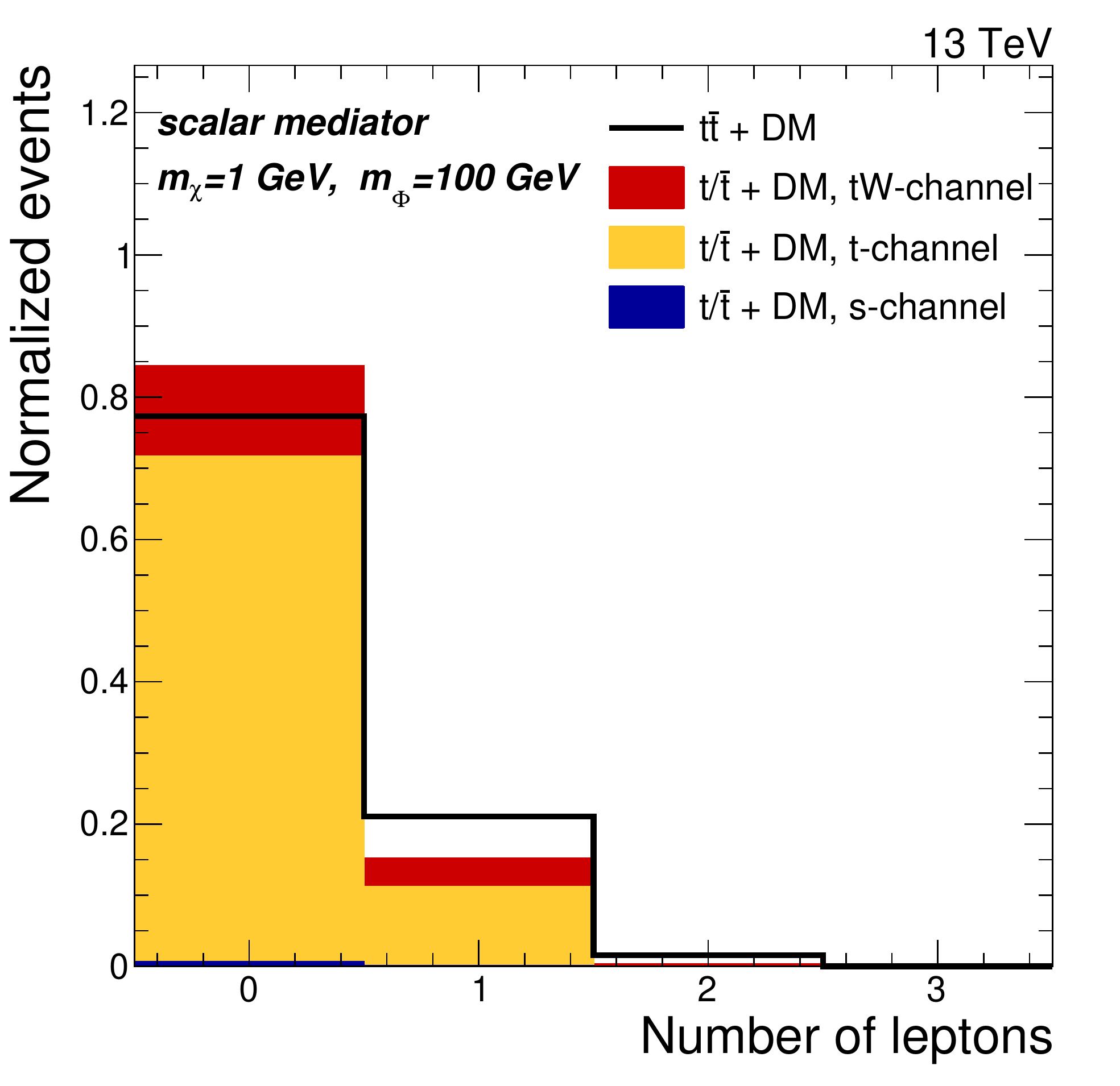}
\par\end{centering}

}
\subfloat[]{\noindent \centering{}\includegraphics[scale=0.26]{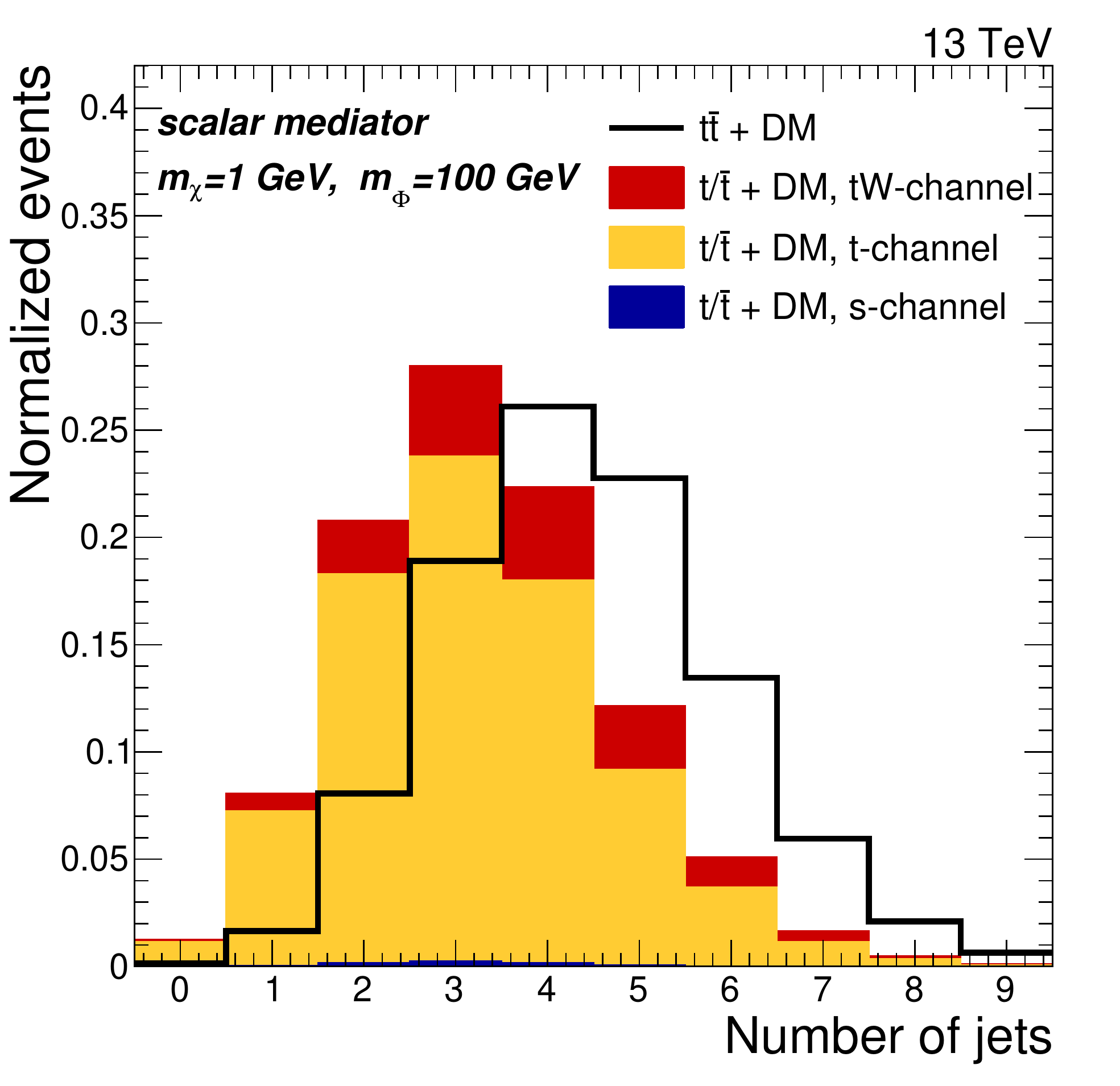}}
\subfloat[]{\noindent \centering{}\includegraphics[scale=0.26]{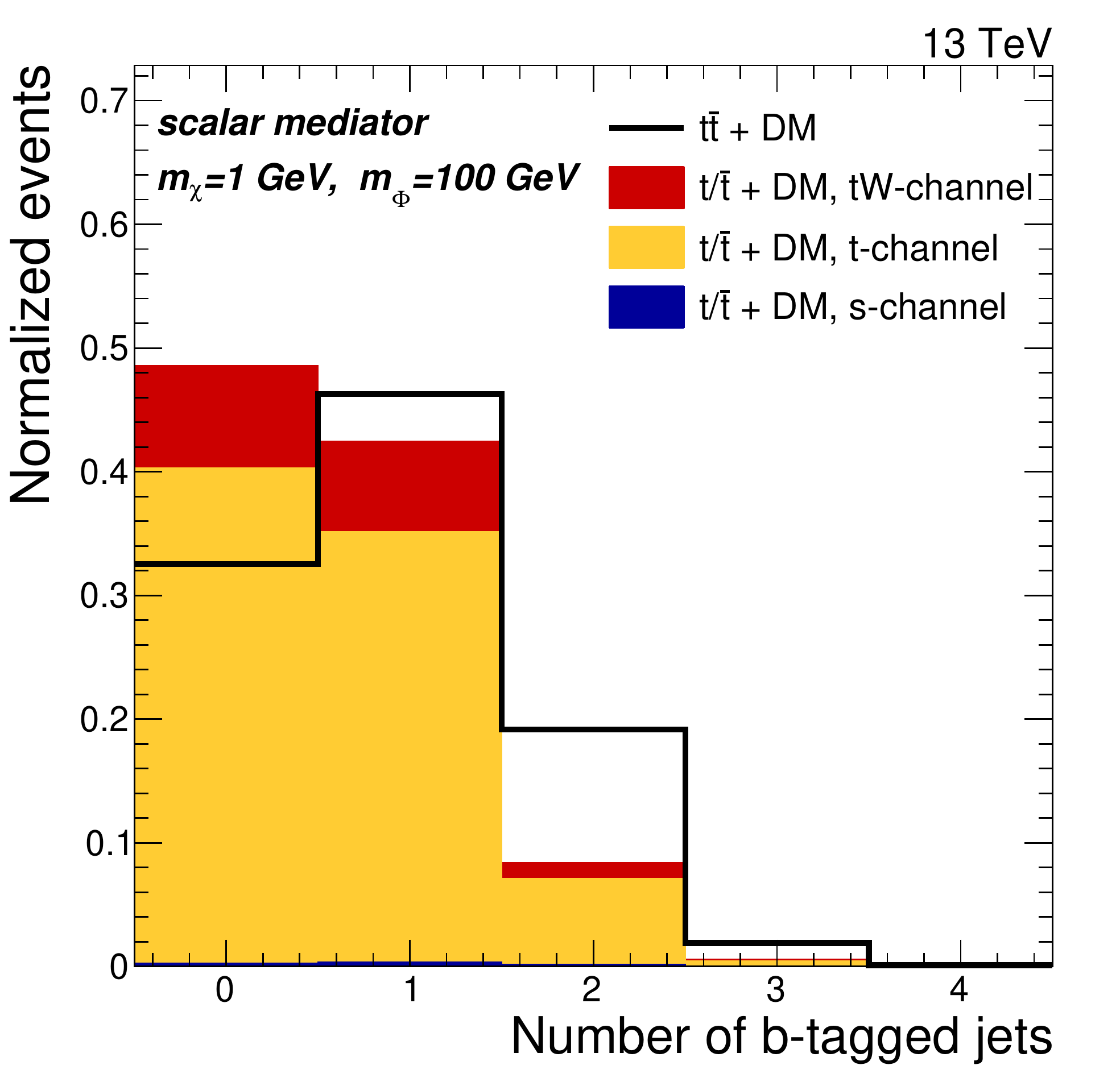}}
\par\end{centering}

\caption{\label{fig:inclusivePlots} Normalized distributions of the number of reconstructed leptons (a), jets (b), and $b$-tagged jets (c) before any selection for a scalar mediator with mass of 100 GeV and a 1 GeV dark matter particle.
The solid black line shows the \ttDM\, process, while the solid histograms are stacked and show the contribution of the \tDM\, processes ($tW$-channel, $t$-channel, and $s$-channel), weighted by the corresponding cross-section. Both \tDM\, and \ttDM\, histograms are normalized to the unit area.}
\end{figure*}

The multiplicity of the reconstructed number of leptons, jets, and $b$-tagged jets are presented in Figure~\ref{fig:inclusivePlots}
for a dark matter particle of mass 1 GeV and a scalar mediator mass of 100 GeV.
These distributions infer that the present selections do not target the optimal phase space for the \tDM\, final state. With respect to \ttDM, the \tDM\, signal has a generally lower jet and $b$-jet multiplicity, and harder \pt\, spectra for the jets and leptons. The former effect is due to the presence of just one genuine $b$ quark in the \tDM\, final state, instead of two $b$ quarks as for the \ttDM\, signal.

The harder \pt\, spectrum is due to the difference between the parton distribution functions relevant to \ttDM\, production as compared to \tDM.
The \ttDM\, production is primarily glue-glue initiated at the LHC, whereas the \tDM\, production modes involve quark p.d.f.s. Compared to gluon p.d.f.s, the quark and antiquark distribution functions drop less rapidly at high parton momentum fraction $x$; as a result the \tDM\, production is less likely to be at threshold. The excess energy allows for a harder \pt\, spectrum than found in the \ttDM\, production.

Given these kinematic differences, a more targeted search could obtain a higher efficiency by lowering the requirements on the number of jets and requiring exactly one $b$-tagged jet in the hadronic channel. A veto on additional $b$-tagged jets could instead reduce the SM $t\bar{t}$ production, which represents the main irreducible background for the considered signal, and further increase the \tDM\, signal efficiency. This phase space is currently not investigated in the \ttDM\, CMS analysis, and can be exploited by a dedicated analysis targeting this production mechanism.

\begin{figure*}
\noindent \begin{centering}
\subfloat[]{\noindent \centering{}\includegraphics[scale=0.26]{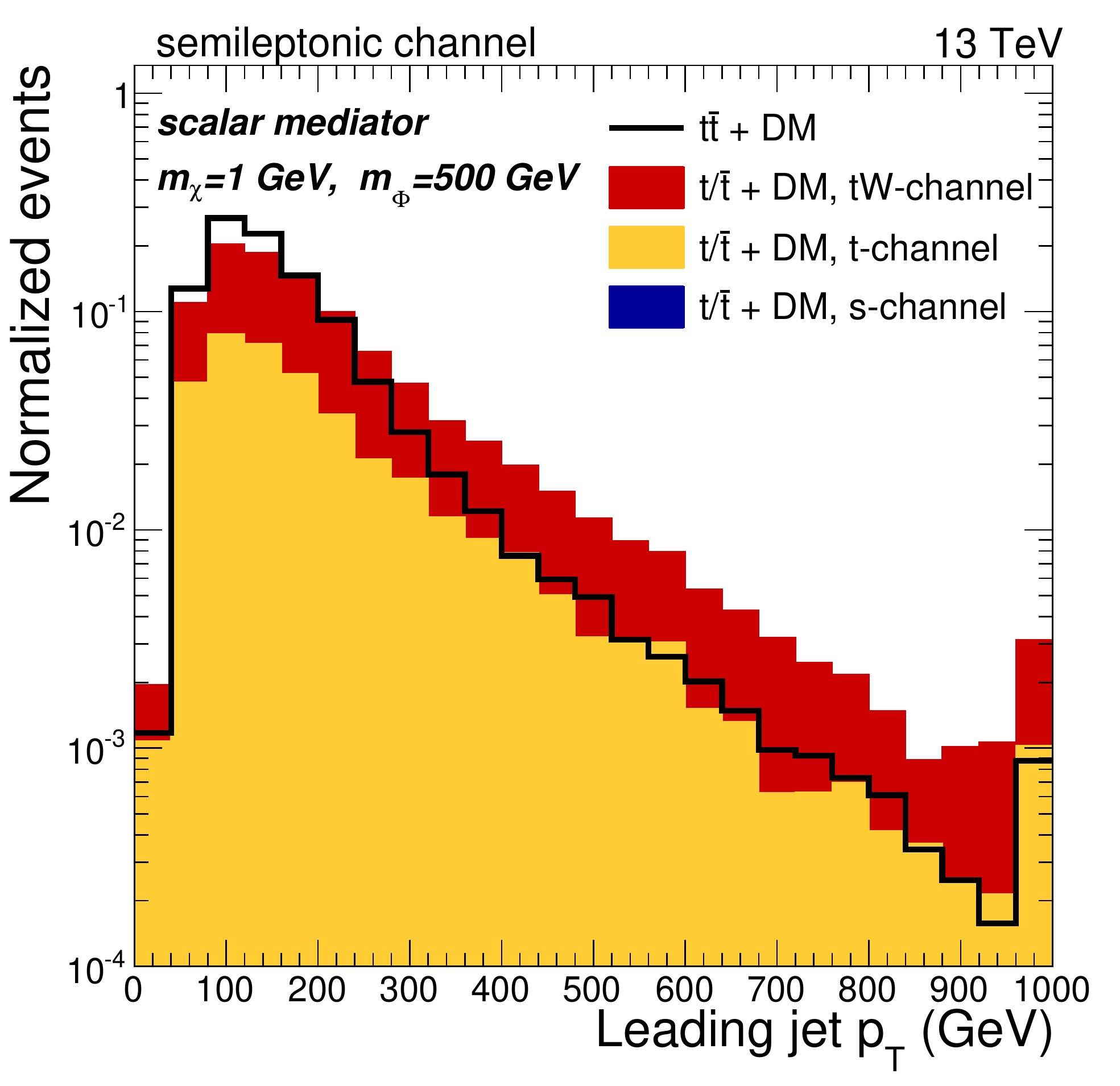}}
\subfloat[]{\noindent \centering{}\includegraphics[scale=0.26]{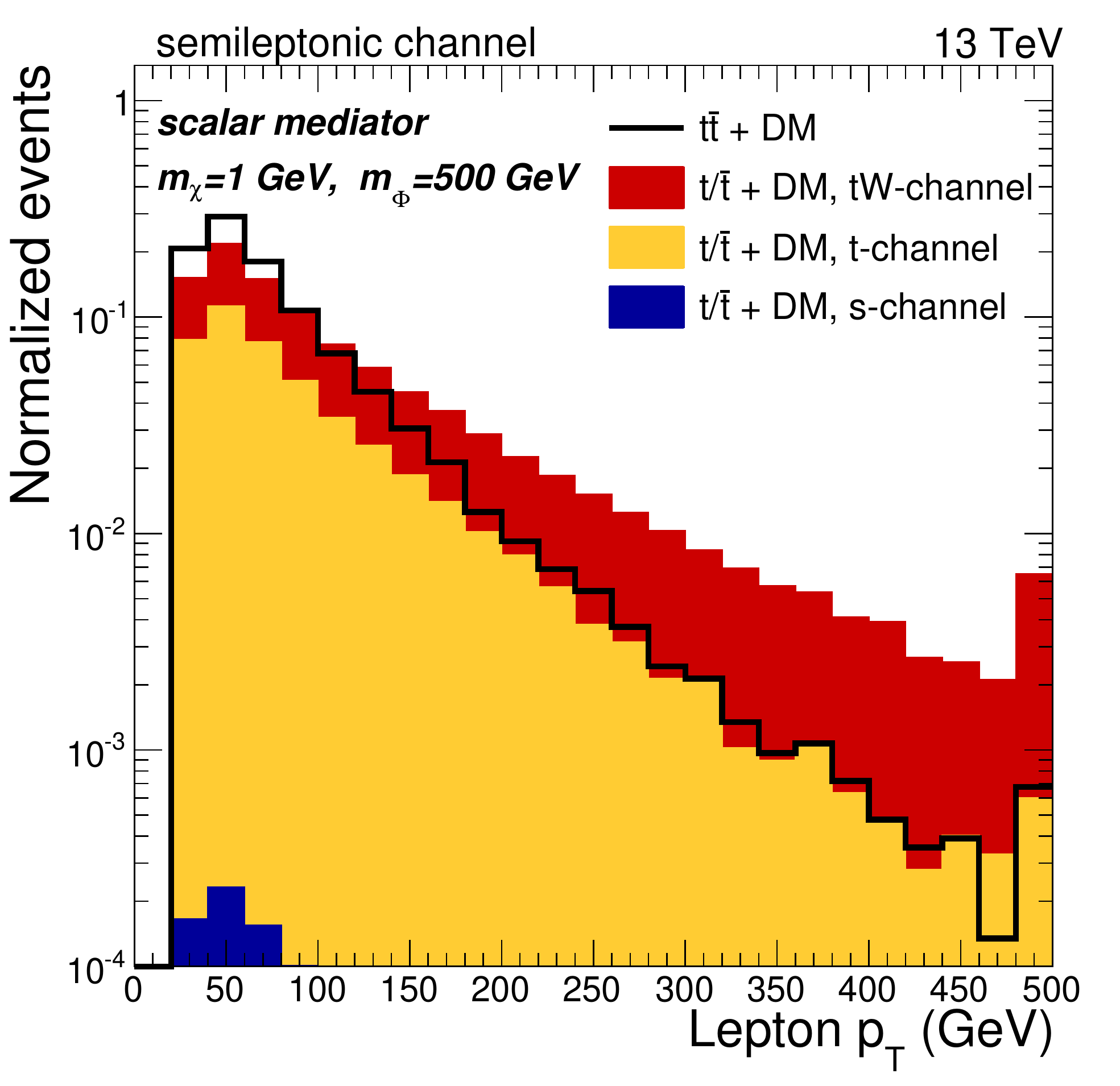}}
\subfloat[]{\noindent \centering{}\includegraphics[scale=0.26]{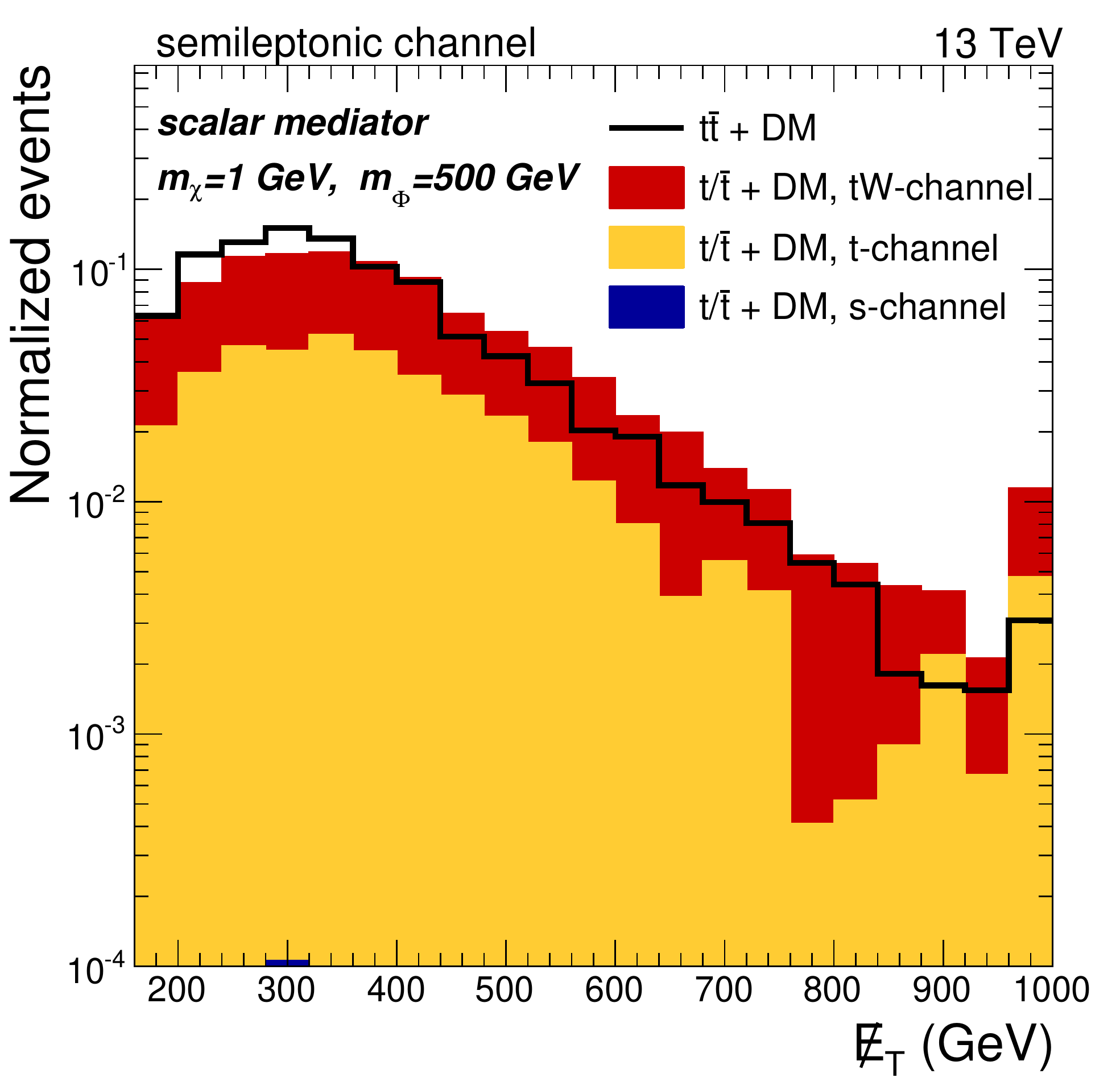}} \\

\subfloat[]{\noindent \centering{}\includegraphics[scale=0.26]{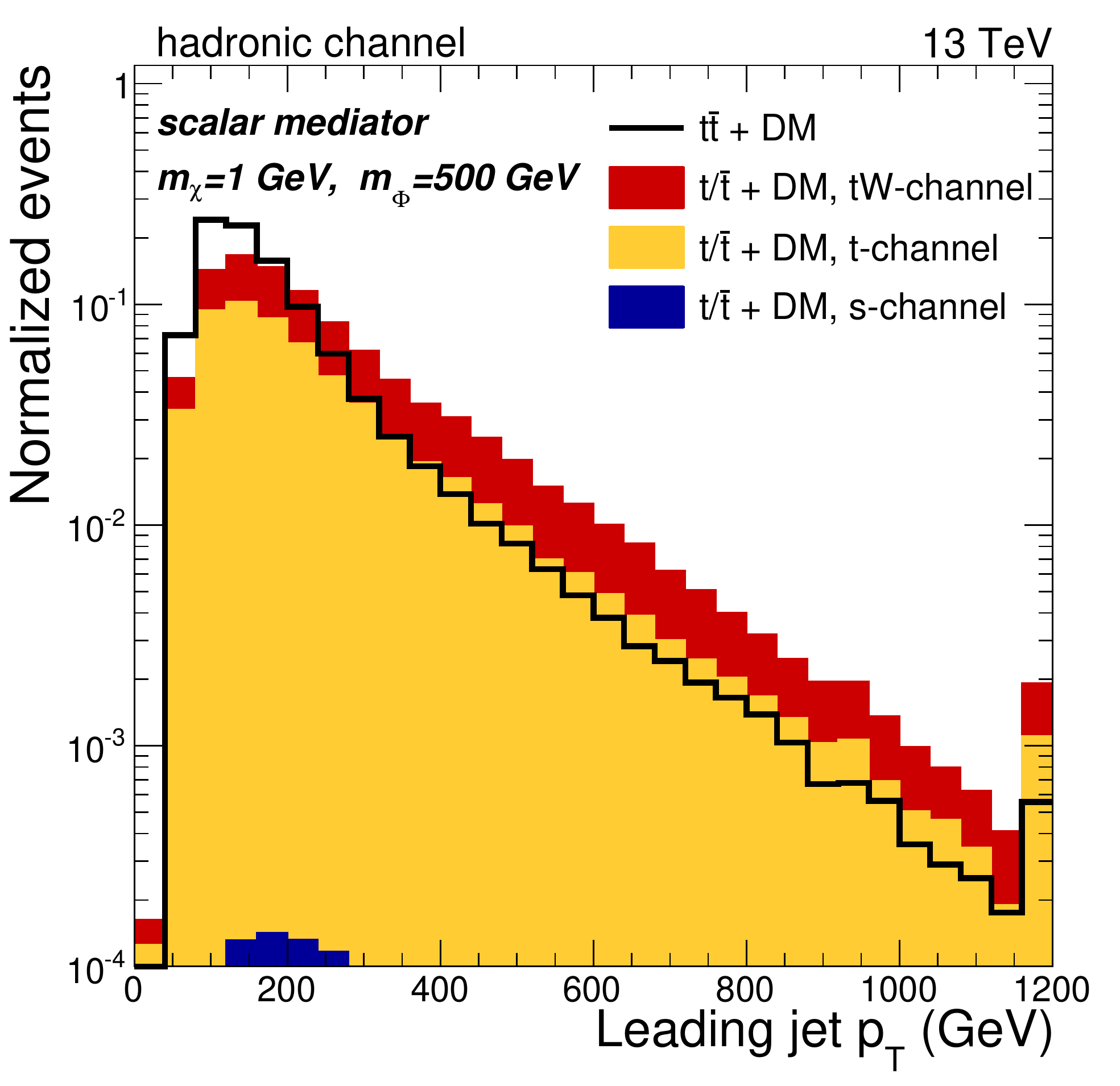}}
\subfloat[]{\noindent \centering{}\includegraphics[scale=0.26]{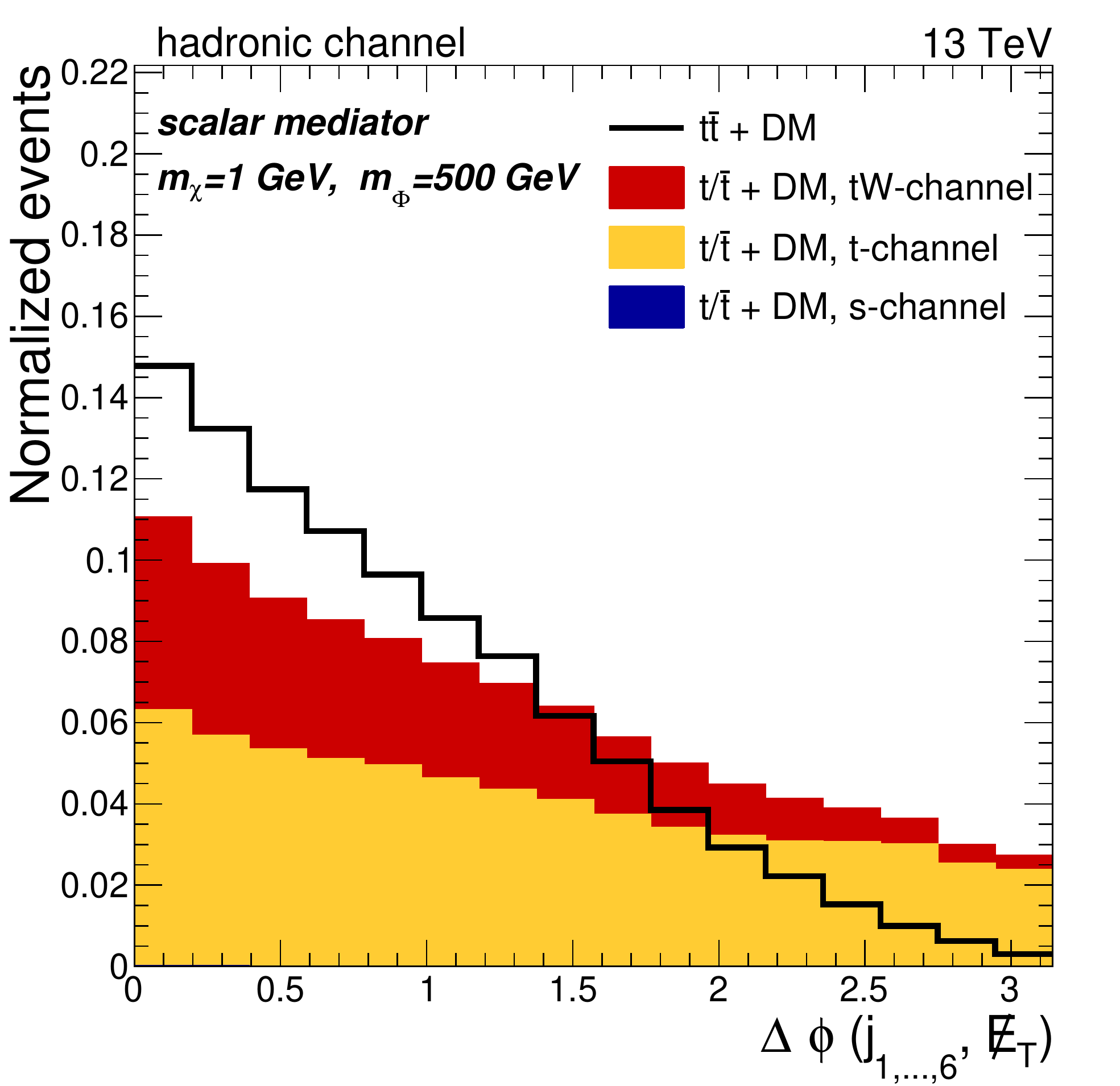}}
\subfloat[]{\noindent \centering{}\includegraphics[scale=0.26]{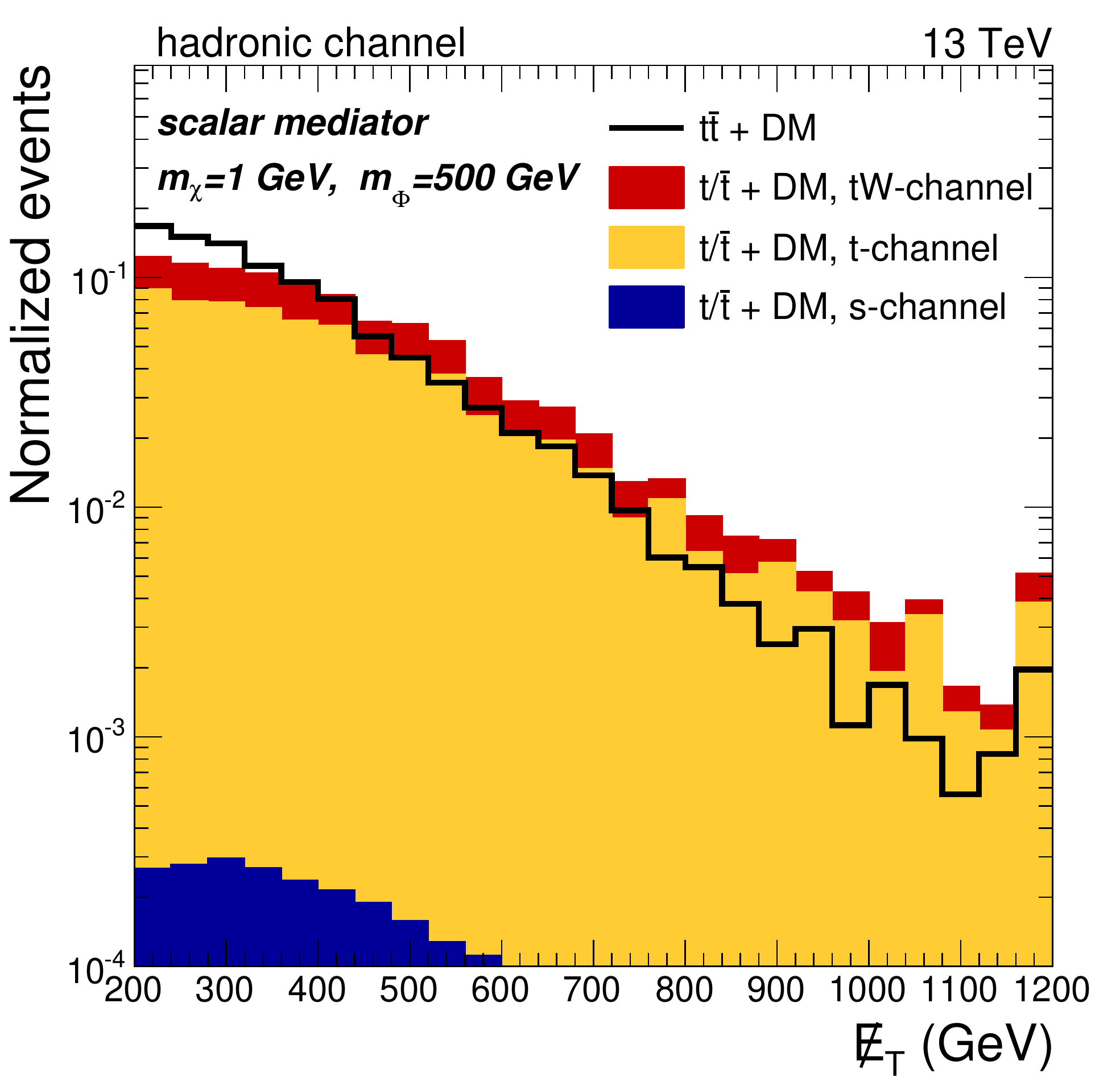}}

\caption{\label{fig:distPlots} Comparison of the \tDM\, (stacked solid histograms) and \ttDM\, kinematic distributions in the semileptonic (top row) and hadronic (bottom row) channels for the \pt\, of the leading jet (left), the $\Delta\phi(j_{i},\MET)$ or the \pt\, of the leading lepton (center), and the reconstructed missing energy (right). The \tDM\, processes ($tW$-channel, $t$-channel, and $s$-channel) are weighted by the corresponding cross section. The \tDM\, and \ttDM\, histograms are normalized to the unit area. The last bin of the distributions includes the overflow events.}
\par\end{centering}
\end{figure*}

The distributions of jets, $b$-jets, leptons and  \MET\, 
are presented in Figure~\ref{fig:distPlots}, showing that \tDM\, 
spectra are harder with respect to \ttDM\, 
 events if the mediator mass is large enough ($m_{\varphi} \gtrsim 100$ GeV).
In fact, despite the generally smaller cross section of the former with respect to the latter, the \tDM\, 
production mode is kinematically favored with respect
to \ttDM~\cite{ReviewPDG}. For lower mediator masses, between 10 and 100 GeV,
the two spectra are comparable.

\section*{Impact on the exclusion limit}

The sensitivity of the \tDM\, process is assessed by calculating the improvement
on the final exclusion limits obtained by the inclusion of \tDM\,  events in addition to the \ttDM\,
process. 
The limits are set on the signal strength $\mu=\sigma/\sigma_{th}$, where $\sigma$ is the measured cross section 
  and $\sigma_{th}$ is the value predicted by the theory model assuming $g_{\chi}=g_{v}=1$.
The expected number of background events and the relative uncertainty are obtained from the CMS search~\cite{CMS-PAS-EXO-16-005} after the background-only fit to data, which provides a reliable estimate of the background yields in the hadronic and semileptonic channels. The signal is normalized to the simulated expectation for both the \tDM\, and \ttDM\, processes. A cut based approach is adopted for events entering the hadronic and semileptonic signal region of the legacy analysis.

The exclusion limits are determined using a modified frequentist approach of confidence levels (CLs)~\cite{CLs,CLs2}, where the asymptotic approximation is taken from the profile likelihood $\mathcal{L}$ as a test statistics~\cite{Asymptotic}. The likelihood function $\mathcal{L}$ is defined as  $\mathcal{L} = \prod_i \mathcal{P} (\lambda_i, n_i)$, and $\mathcal{P} (\lambda, n) = \frac{\lambda^{n} e^{-\lambda}}{n!}$, where $\lambda = \mu \cdot n_{s} + n_{b}$ and $\mu$ is a signal strength modifier, $n_s$ and $n_b$ the number of signal and background events in the $i$-th channel.
Their uncertainties originate from the limited number of generated events, and the background fit to data~\cite{CMS-PAS-EXO-16-005}, respectively, and are treated as nuisance parameters and profiled in the statistical interpretation.
The likelihood is maximized to obtain the best fitted value for $\mu$.

\begin{figure*}\centering
  \includegraphics[width=0.495\textwidth]{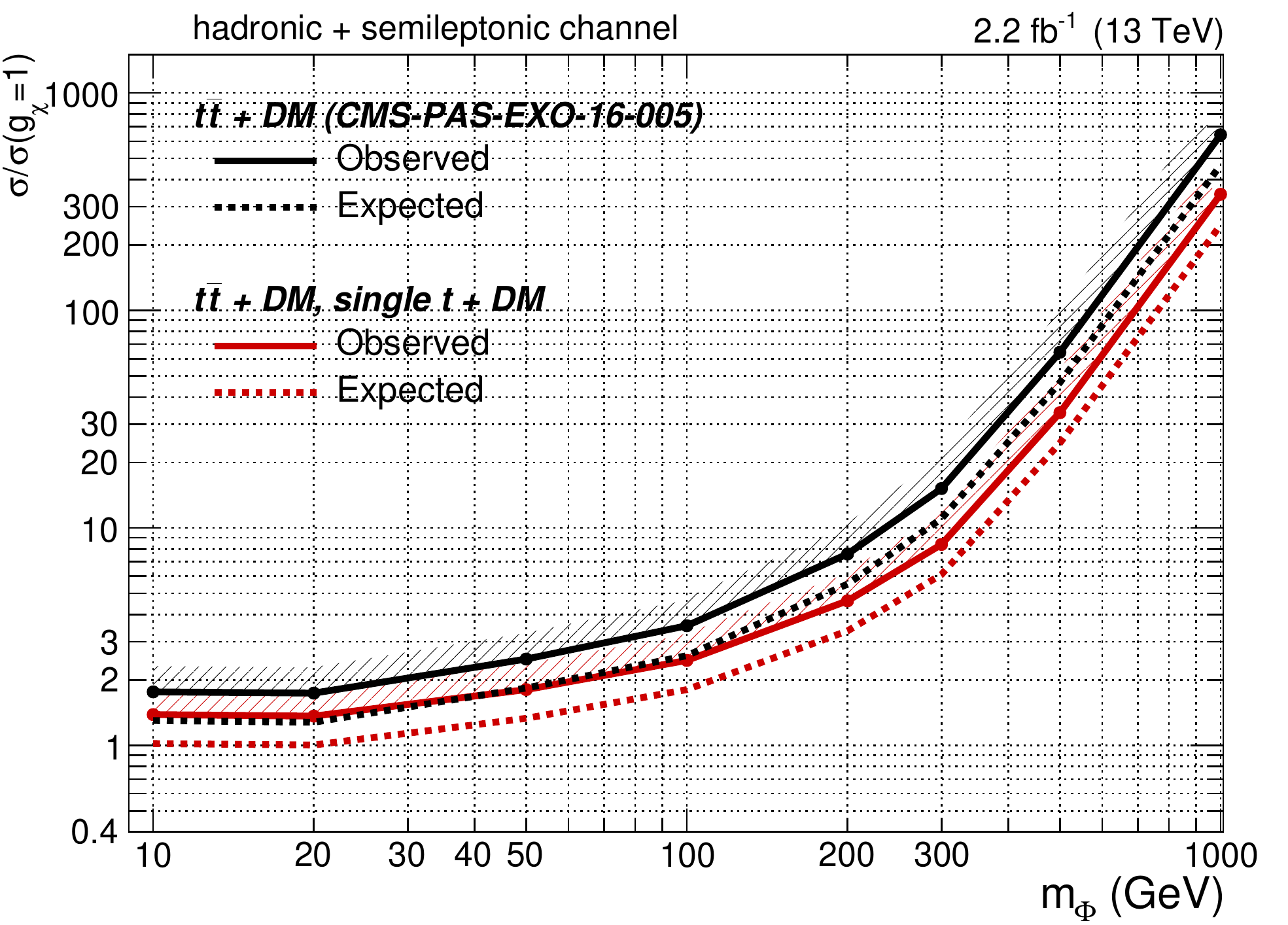}
  \includegraphics[width=0.495\textwidth]{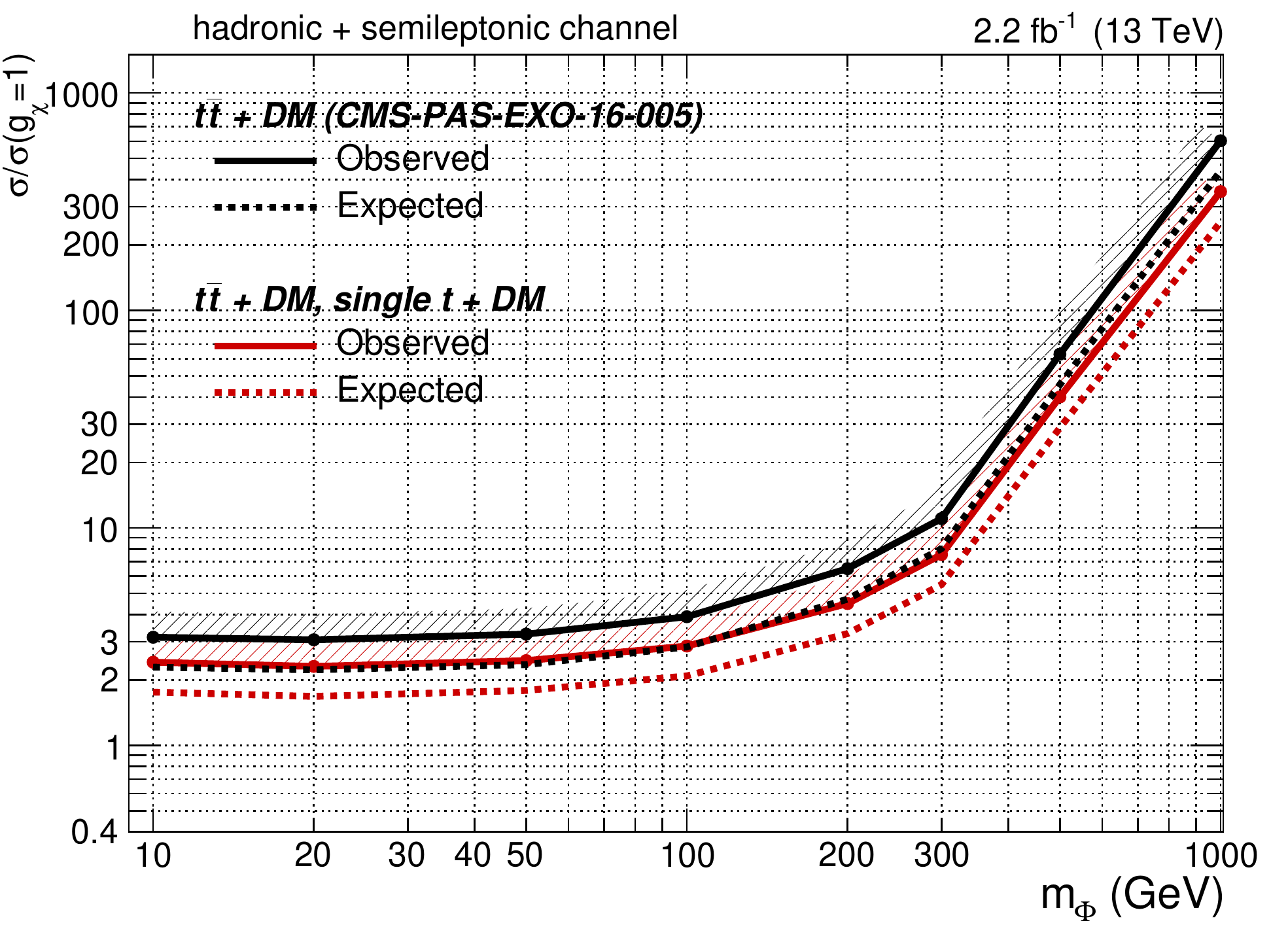}
  \caption{\label{fig:limits} Comparison between the expected and observed exclusion limits on $\mu=\sigma/\sigma_{th}$ considering the \ttDM\, signal alone as in Ref.~\cite{CMS-PAS-EXO-16-005} (black dotted and solid lines), and with the combined \ttDM\, and \tDM\, signals (red dotted and solid lines) for the scalar (left) and pseudoscalar (right) mediator hypothesis. The area above the lines, indicated by the shaded areas, represents the excluded parameter space. The observed number of data events, and the background yields and uncertainties, are taken from Ref.~\cite{CMS-PAS-EXO-16-005}.}
\end{figure*}

The upper limits are reported in Table~\ref{tab:limits} and shown graphically as a function of the mediator mass for the $m_\chi=1$ GeV assumption in Fig.~\ref{fig:limits}, separately for the scalar and pseudoscalar mediators. The resulting values are found to be comparable within 20\% to the inclusive CMS analysis~\cite{CMS-PAS-EXO-16-005} for low mediator mass. At high mediator mass, the CMS search takes advantage of the signal \MET\, shape, which differs significantly from the background, and the exclusion limits derived with the cut-and-count become less stringent than those extracted with the shape fit. Nevertheless, we find that, by including the \tDM\,
processes alongside the \ttDM\, production modes yields relative improvements in the expected limit ranging from $30\%$ at low mediator mass up to $90\%$ at high mediator mass ($m_\varphi=1$ TeV).

\begin{figure*}\centering
  \includegraphics[width=0.495\textwidth]{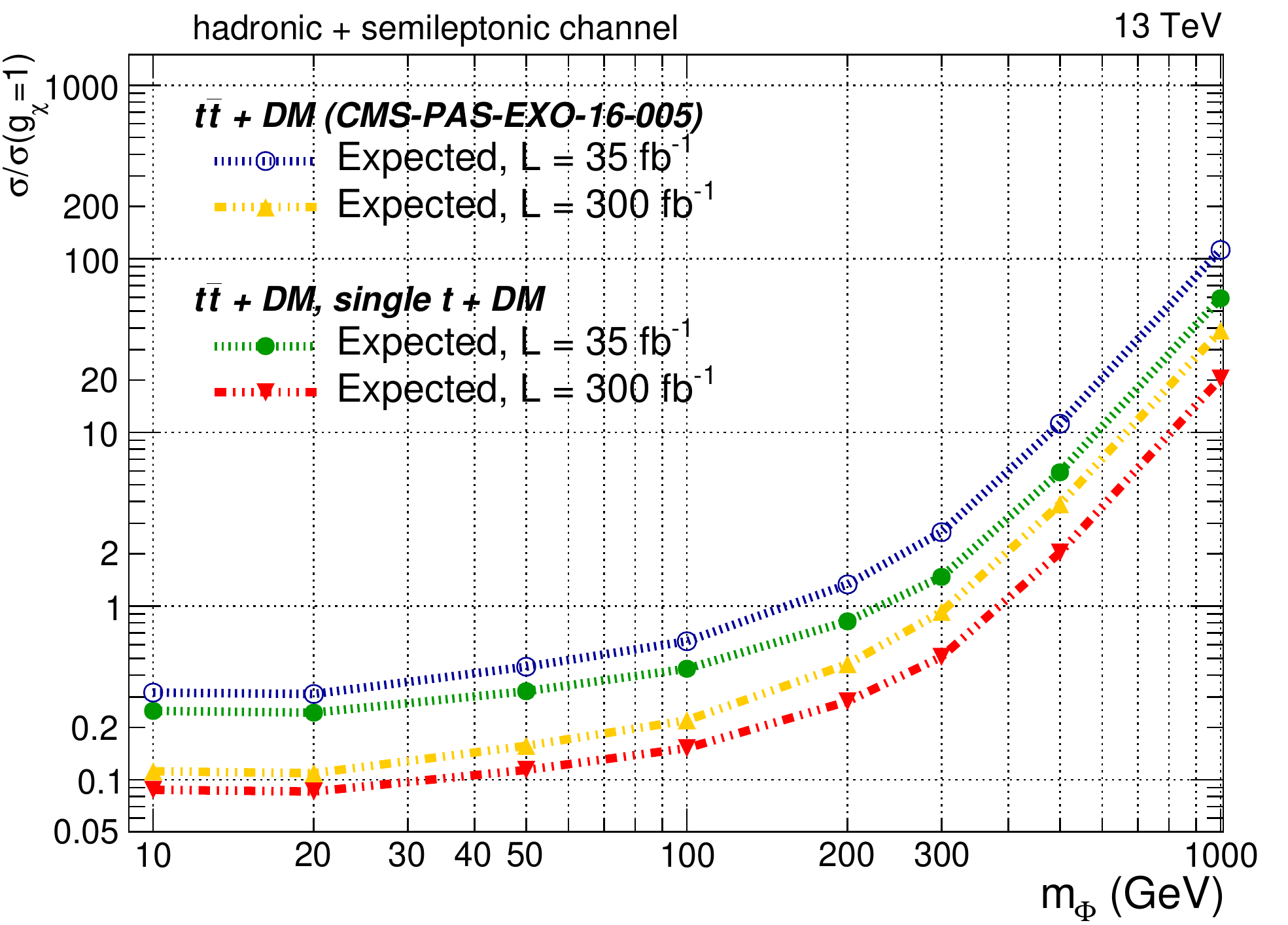}
  \includegraphics[width=0.495\textwidth]{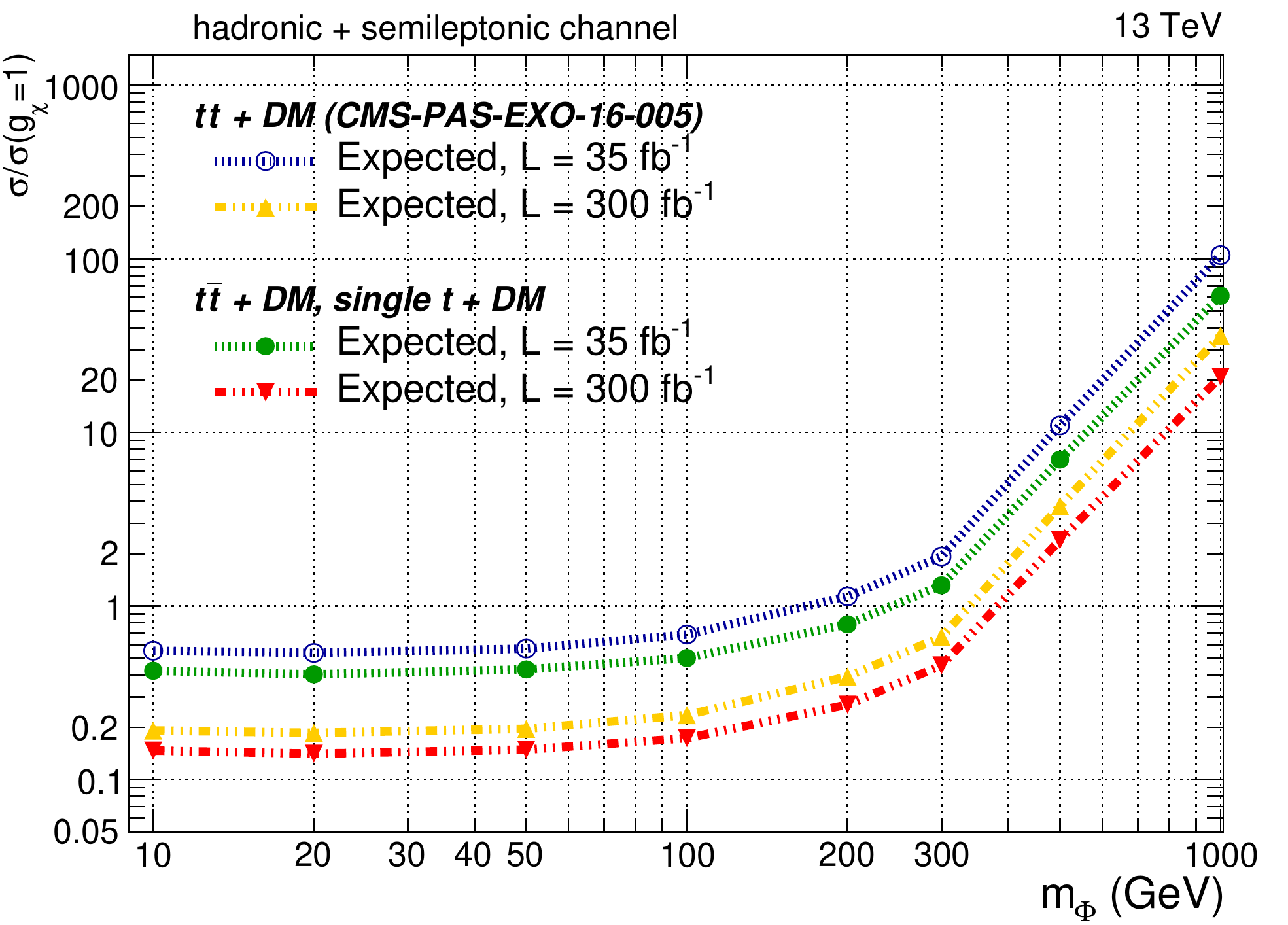}
  \caption{\label{fig:pro} Expected exclusion limits on $\mu=\sigma/\sigma_{th}$ for the \ttDM\, signal (blue and yellow lines) and the sum of \tDM\, and \ttDM\, (green and red lines)f or the scalar (left) and pseudoscalar (right) mediator hypothesis.  The two scenarios considered are with the 2016 dataset ($35 \ \mbox{fb}^{-1}$) and for the LHC Run~II ($300 \ \mbox{fb}^{-1}$).}
\end{figure*}

The sensitivity of the analysis is also projected for two future scenarios. Considering the dataset collected during 2016 and at the end of LHC Run~II, an estimated integrated luminosity of $35 \ \text{fb}^{-1}$ and $\sim 300 \ \text{fb}^{-1}$ will be available to the ATLAS and CMS experiments. Without any reliable method to estimate the systematic uncertainties for these datasets, the results are extracted under the assumption that the background uncertainties will scale as the square root of luminosity and the signal uncertainty remains constant. The exclusion limits are reported in Figure~\ref{fig:pro}.
Even with large luminosities, it is verified that the addition of the \tDM\, signal still increases significantly the sensitivity on the exclusion limit.

\begin{table}[!htb]
\resizebox{\columnwidth}{!}{
\begin{tabular}{cc cc cc}
\toprule 
&\multirow{2}{*}{$m_{\chi},\ m_{\varphi}$ (GeV)} & \multicolumn{2}{c}{\ttDM } & \multicolumn{2}{c}{$t\bar{t},t+$DM }\\
&  &  expected & observed & expected & observed \\
\midrule
\multirow{10}{*}{\rotatebox[origin=c]{90}{scalar}}
  & $m_{\chi}=1,\ m_{\Phi}=10$ & 1.7 & 2.1 & 1.3 & 1.7 \\
 & $m_{\chi}=1,\ m_{\Phi}=20$ & 1.6 & 2.1 & 1.3 & 1.6 \\
 & $m_{\chi}=1,\ m_{\Phi}=50$ & 2.4 & 3.0 & 1.7 & 2.1 \\
 & $m_{\chi}=1,\ m_{\Phi}=100$ & 3.2 & 4.1 & 2.2 & 2.9 \\
 & $m_{\chi}=1,\ m_{\Phi}=200$ & 6.7 & 8.6 & 4.1 & 5.3 \\
 & $m_{\chi}=1,\ m_{\Phi}=300$ & 13 & 17 & 7.4 & 9.5 \\
 & $m_{\chi}=1,\ m_{\Phi}=500$ & 56 & 72 & 29 & 38 \\
 & $m_{\chi}=1,\ m_{\Phi}=1000$ & 554 & 716 & 291 & 377 \\
 & $m_{\chi}=10,\ m_{\Phi}=10$ & 54 & 69 & 37 & 48 \\
 & $m_{\chi}=50,\ m_{\Phi}=300$ & 13 & 17 & 7.7 & 9.8 \\
\midrule
\multirow{10}{*}{\rotatebox[origin=c]{90}{pseudoscalar}}
 & $m_{\chi}=1,\ m_{A}=10$ & 2.8 & 3.6 & 2.1 & 2.8 \\
 & $m_{\chi}=1,\ m_{A}=20$ & 2.7 & 3.5 & 2.1 & 2.6 \\
 & $m_{\chi}=1,\ m_{A}=50$ & 2.9 & 3.7 & 2.2 & 2.8 \\
 & $m_{\chi}=1,\ m_{A}=100$ & 3.4 & 4.4 & 2.5 & 3.2 \\
 & $m_{\chi}=1,\ m_{A}=200$ & 5.7 & 7.3 & 3.9 & 5.0 \\
 & $m_{\chi}=1,\ m_{A}=300$ & 9.6 & 12 & 6.5 & 8.5 \\
 & $m_{\chi}=1,\ m_{A}=500$ & 54 & 70 & 34 & 44 \\
 & $m_{\chi}=1,\ m_{A}=1000$ & 518 & 668 & 299 & 387 \\
 & $m_{\chi}=10,\ m_{A}=10$ & 50 & 65 & 37 & 48 \\
 & $m_{\chi}=50,\ m_{A}=300$ & 9.4 & 12 & 6.4 & 8.3 \\
\bottomrule 
\end{tabular}}\caption{\label{tab:limits} Observed and expected upper limits on $\mu=\sigma/\sigma_{th}$ at 95\% CL, relative to the integrated luminosity collected in 2015 ($L=2.2 \ \mbox{fb}^{-1}$). The center column reports the excluded values for a \ttDM\, signal alone, and the left column for both the \ttDM\, and \tDM\, signal combined.}
\end{table}

\section*{Summary}

We have considered for the first time the production of dark matter in association with a single top quark and found a sizable contribution to dark matter searches with heavy flavour quarks.
Although the cross section is generally smaller with respect to top quark pair plus dark matter processes, the single top quark and dark matter associated production is competitive because of a sizable selection efficiency. Notably, the \pt\, spectra of the visible particles in the event tend to be harder than those found in associated top pair production.

 Including these processes in the search for dark matter produced in association with top quark pairs performed by CMS, an improvement on the exclusion limit between $30$ and $90\%$ is achievable with the 2015 dataset. The increased sensitivity is confirmed also for future analyses with the 2016 dataset and the LHC Run II predicted luminosity. Considering that the present CMS and ATLAS searches do not use any optimized selection for the single top quark and dark matter process, it is reasonable to expect that the reach of single top quark and dark matter will further improve with a dedicated analysis, and it will play a pivotal role in future searches at hadron colliders.

\section*{Acknowledgements}

The work of D. P., A. Z., F. C. is supported in part by the Swiss National Science Foundation (SNF). We would like to thank Bjoern Penning for many helpful discussions related to this paper.


\bibliographystyle{h-physrev5}
\bibliography{DM_meas} 

\end{document}